%% file: main2.tex
\documentclass[reprint,
superscriptaddress,
 amsmath,amssymb,
 aps,
]{revtex4-1}
\usepackage{lineno}
\usepackage{graphicx}
\usepackage{physics}
\usepackage{dcolumn}
\usepackage{bm}
\usepackage{bbold}
\usepackage{color}
\usepackage{hyperref}
\usepackage{qcircuit}
\usepackage{dsfont}

\usepackage{tikz}
\usepackage{pgfplots}
\usetikzlibrary{pgfplots.groupplots}
\usepackage{varwidth}
\usepackage{subfigure}

\newcommand{\beginsupplement}{%
      \setcounter{table}{0}
       \renewcommand{\thetable}{S\arabic{table}}%
        \setcounter{figure}{0}
        \renewcommand{\thefigure}{S\arabic{figure}}%
     }

\begin{document}

\title{Efficient Symmetry-Preserving State Preparation Circuits for the Variational Quantum Eigensolver Algorithm}

\author{Bryan T. Gard}
\email{bgard1@vt.edu}
 \affiliation{%
Department of Physics, Virginia Tech, Blacksburg, VA 24061, U.S.A}%
\author{Linghua Zhu}%
 \affiliation{%
Department of Physics, Virginia Tech, Blacksburg, VA 24061, U.S.A}%
\author{George S. Barron}%
 \affiliation{%
Department of Physics, Virginia Tech, Blacksburg, VA 24061, U.S.A}%
\author{Nicholas J. Mayhall}%
 \affiliation{%
Department of Chemistry, Virginia Tech, Blacksburg, VA 24061, U.S.A}%
\author{Sophia E. Economou}%
 \affiliation{%
Department of Physics, Virginia Tech, Blacksburg, VA 24061, U.S.A}%
\author{Edwin Barnes}%
 \affiliation{%
Department of Physics, Virginia Tech, Blacksburg, VA 24061, U.S.A}%

\date{\today}

\begin{abstract}
The variational quantum eigensolver is one of the most promising approaches for performing chemistry simulations using noisy intermediate-scale quantum (NISQ) processors. The efficiency of this algorithm depends crucially on the ability to prepare multi-qubit trial states on the quantum processor that either include, or at least closely approximate, the actual energy eigenstates of the problem being simulated while avoiding states that have little overlap with them. Symmetries play a central role in determining the best trial states. Here, we present efficient state preparation circuits that respect particle number, total spin, spin projection, and time-reversal symmetries. These circuits contain the minimal number of variational parameters needed to fully span the appropriate symmetry subspace dictated by the chemistry problem while avoiding all irrelevant sectors of Hilbert space. We show how to construct these circuits for arbitrary numbers of orbitals, electrons, and spin quantum numbers, and we provide explicit decompositions and gate counts in terms of standard gate sets in each case. We test our circuits in quantum simulations of the $H_2$ and $LiH$ molecules and find that they outperform standard state preparation methods in terms of both accuracy and circuit depth.
\end{abstract}

\pacs{Valid PACS appear here}
\maketitle


\section{Introduction}
Quantum simulation of fermionic systems, such as molecules, is one of the first envisioned applications of quantum computers, as famously proposed by Feynman~\cite{Feynman1982}. The first protocol introduced for quantum chemistry simulations is based on the so-called phase estimation algorithm~\cite{Abrams1999,Aspuru-Guzik2005}. This algorithm however requires a large number of quantum gates, leading to long quantum circuits that are challenging for existing and near-term noisy intermediate scale quantum (NISQ) devices \cite{Preskill2018,Kuhn2018}. For such devices, alternative, hybrid algorithms are instead envisioned. In such algorithms, the work is shared between a quantum processor and a classical computer. In particular, the variational quantum eigensolver (VQE), first introduced and demonstrated experimentally by Peruzzo et al.~\cite{Peruzzo2014}, has become the prevailing algorithm for chemistry simulations with NISQ devices, with several milestone papers demonstrating the calculation of molecular energies and wavefunctions~\cite{OMalley2016, Kandala2017, Colless2018, Shen2017,Hempel2018,Yunseong2019}.

The VQE algorithm relies on preparing and measuring multi-qubit states based on a variational ansatz, and using the classical computer to optimize and update the variational parameters in this ansatz. Some of the advantages of VQE are that its variational character can provide some degree of error mitigation in the gates~\cite{OMalley2016,McArdle2018,Bonet2018,Barkoutsos2018}, and that it features shallower circuits compared to the phase estimation algorithm.
The form of the ansatz is a crucial ingredient of VQE and one that can determine its success on NISQ devices. There are two main approaches in determining the ansatz. One approach is based on a technique from chemistry, the unitary coupled cluster method~\cite{BARTLETT1989, Kutzelnigg1991,Ryabinkin2018}, translated into quantum gates by Trotterization~\cite{Ryabinkin2018}. This approach tends to lead to deeper circuits than what is currently feasible on hardware, and is generally not exact. To address these issues, a new, iterative algorithm termed ADAPT-VQE, was recently put forward and was shown to enable a much more compact ansatz while simultaneously exhibiting higher accuracy~\cite{Grimsley2018}. An alternative approach is to base the ansatz on the capabilities of the hardware and prepare states by combining parameterized gates available on the processor~\cite{Kandala2017,Bian2019}. Such an ansatz has the advantage of compatibility with the capabilities of the hardware, and as such is NISQ-friendly. On the other hand, in its simplest form it is an ad hoc ansatz that can cause the algorithm to get stuck on ‘barren plateaus’~\cite{McClean2018} as the number of qubits increases and the Hilbert space correspondingly grows exponentially. Therefore, for hardware-based ans\"atze to be a viable approach for problems of interest, they must be selected in a way that guarantees they span the part of the Hilbert space where the solution lives, while avoiding generating unphysical states.

Two ways to guarantee that the desired part of the Hilbert space is accessed include adding terms in the VQE energy function that penalize symmetry violations \cite{McClean2016,Ryabinkin2019} or carefully designing state preparation circuits so that they only produce states with the appropriate symmetries regardless of how their variational parameters are chosen. An early step toward the latter direction was taken by Wang et al. \cite{Wang2009}, who focused on the preparation of states with a well-defined number of occupied spin orbitals and showed that the number of CNOT gates required for this scales polynomially with the number of qubits in the limit where the number of electrons is much smaller than the number of qubits. More recently, Barkoutsos et al.~\cite{Barkoutsos2018} enforced particle number conservation by using the particle-hole representation in conjunction with a parametrized particle-conserving exchange-type gate~\cite{Roth2017,Egger2019,Sagastizabal2019,Ganzhorn2018}, which we also make use of here. However, important open questions remain, including how other symmetries can also be built into the circuits and whether more efficient circuits containing the minimal number of parameters necessary to span the symmetry subspace exist.

In this paper, we address these questions by introducing state preparation circuits that provide a systematic, economical way to generate states with well-defined symmetries, including particle number, total spin, spin projection, and time-reversal. Our circuits incorporate the minimal number of parameters needed to \emph{fully} span the appropriate symmetry subspace while avoiding all states outside this subspace.This general approach has two key advantages: the first is that the true ground state is guaranteed to be contained within the space of states spanned by the circuit, and the second is that resources are not spent on generating irrelevant parts of the Hilbert space, reducing the complexity of the classical optimization step of the VQE algorithm. Eliminating extraneous parameters can dramatically speed up the optimization process and suppress the probability of getting trapped in local extrema or barren plateaus. We present circuits for arbitrary numbers of single-particle orbitals and electrons and for arbitrary spin quantum numbers. Our circuits are constructed with hardware constraint considerations, including a reduced number of CNOT gates that need only be applied between adjacent qubits in a linear array, making our work particularly suited for NISQ devices. In addition, since our construction conserves number and spin symmetries, symmetry verification techniques can be used to mitigate any errors which violate these symmetries~\cite{Bonet2018,McArdle2018,Sagastizabal2019}. 
Our most general circuits, which conserve particle number, time-reversal, and spin symmetries, are guaranteed to span the appropriate symmetry subspace by construction. We also present circuits that conserve particle number and time-reversal symmetries, but not total spin; the symmetry-preserving properties of these circuits are verified with numerical calculations. We test the performance of our circuits against standard state preparation ans\"atze by running VQE simulations of the $H_2$ and $LiH$ molecules. We find that our circuits outperform the standard methods in terms of both accuracy and circuit depth.

The paper is organized as follows. In Sec.~\ref{sec:number}, we define the basic gate set we use to manipulate particle number on the quantum processor and show how these gates can be systematically assembled into circuits that preserve particle number symmetry. We also discuss how to respect time-reversal symmetry as well. In Sec.~\ref{sec:spin}, we show how to create circuits that respect total spin and spin projection symmetries in addition to particle number. We give some concluding remarks in Sec.~\ref{sec:conclusion}. An appendix contains additional details about gate decompositions.

\section{Results}\label{sec:number}
\subsection{Particle number and time-reversal symmetries}

In this work, we focus on mapping chemistry problems onto  quantum processors using the Jordan-Wigner mapping \cite{Aspuru-Guzik2005,Jordan1928}, in which each qubit in the quantum processor corresponds to a particular spin-orbital, and the qubit states $\ket{0}$ and $\ket{1}$ encode the occupation of that spin-orbital. Any multi-electron state involving $n$ spin-orbitals on the chemistry side can be mapped to a corresponding state of $n$ qubits on the quantum processor. In this mapping, fixing the total number of electrons is tantamount to fixing the total number of qubits that are in the excited state $\ket{1}$. Thus, the Jordan-Wigner mapping relates fixed-particle-number subspaces to fixed-excitation subspaces in the qubit Hilbert space.

Formally, we can define the qubit subspace corresponding to $m$ electrons occupying $n$ spin-orbitals as
\begin{equation}
    H_{n,m} = \text{span}\left\{ \ket{s_1,s_2,...,s_n} \bigg|s_i\in\{0,1\}, \sum_{i=1}^{n} s_i = m \right\}.
\end{equation}
This is the subspace of multi-qubit states containing $m$ qubits in the $\ket{1}$ state and $n-m$ qubits in the $\ket{0}$ state, so that $\text{dim}(H_{n,m}) =\binom{n}{m}$. A general state in this subspace can be represented as an arbitrary superposition of these basis states with complex coefficients and is thus characterized by $2~\text{dim}(H_{n,m})-2$ real parameters, where we have removed two parameters by fixing the normalization and neglecting a global phase. In the absence of any other symmetries, $2~\text{dim}(H_{n,m})-2$ is the minimal number of real variational parameters needed to prepare arbitrary trial states describing the correct number of electrons.

An additional symmetry that often arises in chemistry problems is time-reversal symmetry. This symmetry is typically present, for example, when one wants to solve the stationary Schr\"odinger equation in the absence of any applied magnetic field. In this case, one can always choose the energy eigenstates to be strictly real functions. Under the Jordan-Wigner mapping, this means that the coefficients appearing in the multi-qubit superposition states we prepare as trial states for the VQE algorithm should be restricted to real values. This will reduce the dimensionality of the target symmetry subspace by a factor of two down to
$\text{dim}(H_{n,m})-1$. Imposing this restriction on the trial states will prevent the classical optimizer from wasting time exploring a large portion of Hilbert space that does not contain any of the desired energy eigenstates. When we introduce our state preparation circuits below, we will see that time-reversal symmetry can be imposed easily after other symmetries are already built into the circuits, essentially just by fixing half of the variational parameters in the circuits in such a way that the resulting states are strictly real.

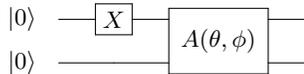
\begin{figure}[!tb]
\[ \Qcircuit @C=1.5em @R=.7em {
\ket{0} &	&	\gate{X}	&	\multigate{1}{A(\theta,\phi)}	&	\qw	\\
\ket{0} &	&	\qw	&	\ghost{A(\theta,\phi)}	&	\qw
} \]
\caption{A simple example of a 2 qubit circuit which exactly spans the subspace defined by 1 excitation, $\alpha\ket{01} + \beta\ket{10}$, with two parameters $(\theta,\phi)$. This circuit saturates the lower bound on the number of real parameters required to construct arbitrary states in $H_{2,1}$.}
\label{fig:2q1e}
\end{figure}

Before we introduce our general scheme for constructing particle-number-conserving state preparation circuits, we first present a few simple examples that may provide some intuition about the general structure of such circuits. First note that the cases $m=0$ and $m=n$ are trivial since in each case, there is only a single state spanning the subspace. Therefore, we restrict our attention to $0<m<n$ throughout this work. The simplest nontrivial example is the case of $m=1$ electron in $n=2$ orbitals. (Since we are presently only concerned with particle-number symmetry, these could be spin-orbitals or spatial orbitals in the case of spinless fermions. Spin symmetries will be incorporated in the next section.) A circuit that spans the corresponding subspace $H_{2,1}$ is shown in Fig.~\ref{fig:2q1e}.

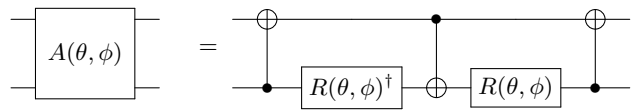
\begin{figure}[!htb]
\[ \Qcircuit @C=1em @R=0.75em {
&	\multigate{2}{A(\theta,\phi)}	&	\qw	&		&		&		&	\targ	&	\qw	&	\ctrl{+2}	&	\qw	&	\targ	&	\qw	\\
&		&		&		&	=	&		&		&		&		&		&		&		\\
&	\ghost{A(\theta,\phi)}	&	\qw	&		&		&		&	\ctrl{-2}	&	\gate{R(\theta,\phi)^\dagger}	&	\targ	&	\gate{R(\theta,\phi)}	&	\ctrl{-2}	&	\qw
} \]
\caption{Decomposition of the $A$ gate in terms of elementary single and two-qubit gates. $R(\theta,\phi)=R_z(\phi+\pi)R_y(\theta+\pi/2)$, where $R_{z}(\theta )=\exp (-i\theta \sigma _{z}/2)$, $R_{y}(\phi )=\exp (-i\phi \sigma _{y}/2)$.}
\label{fig:Adecomp}
\end{figure}

This circuit requires only $2$ parameters to span the single-excitation (one-electron) subspace, which is comprised of states of the form $\alpha \ket{01} + \beta \ket{10}$. Here, although $\alpha$ and $\beta$ are complex, they contribute only $2$ real parameters after we impose normalization $|\alpha|^2+|\beta|^2=1$ and discard a global phase. The key ingredient in this circuit is a two-qubit entangling gate that we have denoted as $A(\theta,\phi)$. In the basis $\ket{00}$, $\ket{01}$, $\ket{10}$, $\ket{11}$, it is defined as \cite{Barkoutsos2018}
\begin{eqnarray}
A(\theta ,\phi )=\begin{pmatrix}
1 & 0 & 0 & 0\\
0 & \cos \theta  & e^{i\phi }\sin \theta  & 0\\
0 & e^{-i\phi }\sin \theta & -\cos \theta & 0\\
0 & 0 & 0 &1
\end{pmatrix}.\label{eq:Agate}
\end{eqnarray}
It is clear from the form of this exchange-type gate that it preserves particle number since it mixes $\ket{01}$ and $\ket{10}$ but does nothing to the $\ket{00},\ket{11}$ subspace. The initial $X$ gate on the first qubit in Fig.~\ref{fig:2q1e} brings the two-qubit state into the one-excitation subspace, while the subsequent $A$ gate generates all possible superpositions within this subspace, as can be seen by inspection in this case. If we wish to also impose time-reversal symmetry, then it suffices to set $\phi=0$ in each of the $A$ gates. This removes the phase from each coefficient in the resulting superposition state without restricting the magnitude, thus ensuring that the resulting state is an arbitrary real state. The $A$ gate plays a central role in our state preparation circuits. It can be decomposed into a sequence of two single-qubit gates and three CNOT gates, as shown in Fig.~\ref{fig:Adecomp}. This decomposition is minimal in the number of CNOT gates. In operator form, the $A$ gate is
\[A=e^{-\frac{i}{2}(\frac{\pi}{2}-\phi) Z_2}e^{-\frac{i}{2}(\theta X_1 X_2+\theta Y_1 Y_2+ \frac{\pi}{2} Z_1 Z_2)}e^{-\frac{i}{2}(\frac{\pi}{2}Z_1-\phi Z_2)},
\]
up to an irrelevant global phase \cite{Zhang2003}.

Building upon the example of Fig.~\ref{fig:2q1e}, we find that we can generate trial states corresponding to other particle and orbital numbers using a similar construction.
\begin{figure}[!tb]
\[ \Qcircuit @C=0.5em @R=.7em {
\ket{0} &	&	\qw	&	\multigate{1}{A(\theta_1,\phi_1)}	&	\qw	&	\multigate{1}{A(\theta_4,\phi_4)}	&	\qw	&	\qw	\\
\ket{0} &	&	\gate{X}	&	\ghost{A(\theta_1,\phi_1)}	&	\multigate{1}{A(\theta_3,\phi_1)}	&	\ghost{A(\theta_4,\phi_4)}	&	\multigate{1}{A(\theta_6,\phi_6)}	&	\qw	\\
\ket{0} &	&	\gate{X}	&	\multigate{1}{A(\theta_2,\phi_1)}	&	\ghost{A(\theta_3,\phi_1)}	&	\multigate{1}{A(\theta_5,\phi_5)}	&	\ghost{A(\theta_6,\phi_6)}	&	\qw	\\
\ket{0} &	&	\qw	&	\ghost{A(\theta_2,\phi_1)}	&	\qw	&	\ghost{A(\theta_2,\phi_2)}	&	\qw	&	\qw
} \]
\caption{An example circuit for the case of $n=4,m=2$ which exactly spans the subspace defined by six basis states using the minimal number (10) of parameters. Note that $\phi_1$ is used three times.}
\label{fig:4q2e}
\end{figure}
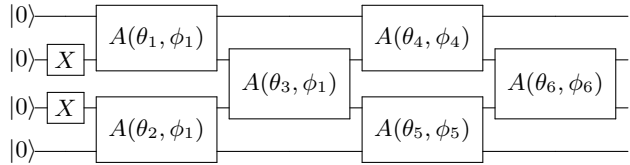
\begin{figure}[!tb]
\[ \Qcircuit @C=1.25em @R=.7em {
\ket{0} &	&	\gate{R(\alpha,\beta)}	&	\ctrl{+1}	&	\gate{R(\gamma,\delta)}	&	\ctrl{+2}	&	\qw	&	\qw	&	\multigate{1}{A(\xi,\chi)}	&	\qw	\\
\ket{0} &	&	\qw	&	\targ	&	\gate{R(\epsilon,\zeta)}	&	\qw	&	\ctrl{+2}	&	\qw	&	\ghost{A(\xi,\chi)}	&	\qw	\\
\ket{0} &	&	\qw	&	\qw	&	\qw	&	\targ	&	\qw	&	\gate{X}	&	\multigate{1}{A(\theta,\phi)}	&	\qw	\\
\ket{0} &	&	\qw	&	\qw	&	\qw	&	\qw	&	\targ	&	\gate{X}	&	\ghost{A(\theta,\phi)}	&	\qw	
} \]
\caption{Another example circuit for the case of $n=4,m=2$, which also spans the desired subspace with the minimal 10 parameters, but with fewer two-qubit gates than the circuit shown in Fig.~\ref{fig:4q2e}. The single-qubit gates $R(\theta,\phi)$ are as defined in Fig.~\ref{fig:Adecomp}.}
\label{fig:4q2ev2}
\end{figure}
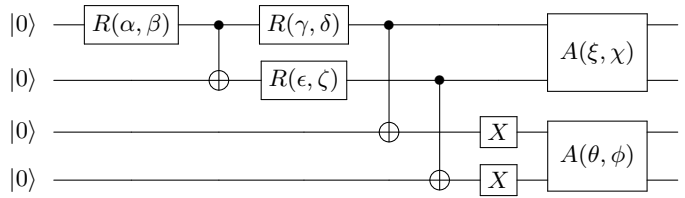
Fig.~\ref{fig:4q2e} shows another example circuit for the case of $m=2$ electrons in $n=4$ orbitals. In this case, we start with two $X$ gates to bring the state into the two-excitation subspace, and we then apply a series of $A$ gates between neighboring qubits to create arbitrary superpositions within this subspace. In this example, an arbitrary complex two-electron state is specified by $2\binom{4}{2}-2=10$ real parameters. Since each $A$ gate introduces two parameters, one might expect that only five $A$ gates would be needed to generate an arbitrary state. However, we find that at least six $A$ gates are needed to do this, although two of the parameters can be fixed to reduce the total parameter count back down to 10. We can again impose time-reversal symmetry by setting all the $\phi_i$ parameters to zero in the $A$ gates, which reduces the parameter count down to 6---one more than the minimal number of 5. To remove this extra parameter, we find numerically that it works to set either $\theta_4$ or $\theta_5$ to zero. Note that setting both $\theta_i$ and $\phi_i$ to zero does not remove the $i$th $A$ gate completely but instead reduces it to a $CZ$ gate, as is evident from Eq.~\eqref{eq:Agate}. 

To confirm that this and all other circuits presented in this section indeed span the target subspace, we compute the fidelity $F=\frac{1}{N}\sum_{i=1}^{N}|\langle\Psi_i|\psi_i\rangle|^2$, where $\ket{\Psi_i}$ is a random state within the chosen subspace, and $\ket{\psi_i}$ is the state output by our circuit after we maximize $|\langle\Psi_i|\psi_i\rangle|^2$ with respect to the variational parameters. We check numerically that $F=1$ can be achieved using the minimal number of parameters, $2~\textrm{dim}(H_{n,m})-2$. We choose enough random states, $N \gg \textrm{dim}(H_{n,m})$, to ensure that the subspace is adequately represented. This is how we determine that, in the case of Fig.~\ref{fig:4q2e} with time-reversal symmetry, it works to set $\theta_4$ or $\theta_5$ to zero, but setting other $\theta$ parameters to zero instead does not achieve unit fidelity (although the fidelity still remains very high). Further details about the numerical verification of our circuits are given in the supplementary information~\cite{GardSup2019}.

Although we have found a circuit that prepares all states in the two-excitation subspace using the minimal number of variational parameters (Fig.~\ref{fig:4q2e}), this solution is neither unique nor optimal in terms of the number of CNOT gates. To illustrate these points, we present another circuit that accomplishes the same task in Fig.~\ref{fig:4q2ev2}. In addition to the $A$ gate, this circuit also makes use of single-qubit gates beyond just $X$ gates. This example is quite different from the one shown in Fig.~\ref{fig:4q2e} in that it does not first apply $X$ gates to two qubits in order to bring the quantum processor into the appropriate particle-number subspace. Instead, the appropriate subspace is approached gradually as the circuit is performed, making it more challenging to understand and generalize the circuit. Since each $A$ requires three CNOT gates to implement, we see that the circuit in Fig.~\ref{fig:4q2ev2} requires only 9 CNOT gates, while the one in Fig.~\ref{fig:4q2e} requires twice as many. However, it should be noted that the circuit of Fig.~\ref{fig:4q2e} requires only nearest-neighbor qubit coupling, which is not true of the one in Fig.~\ref{fig:4q2ev2}. This example highlights the fact that further reductions in circuit depth are possible even if the circuit contains the minimal number of parameters, although this may require an increase in the qubit connectivity.

\begin{figure}[!tb]
\[ \Qcircuit @C=1.25em @R=1em {
\ket{0} &	&	\qw	&	\multigate{1}{A}	&	\qw	&	\qw	&		&		&		&	\qw	&	\multigate{1}{A}	&	\qw	&	\qw	\\
\ket{0} &	&	\gate{X}	&	\ghost{A}	&	\multigate{1}{A}	&	\qw	&		&		&		&	\qw	&	\ghost{A}	&	\multigate{1}{A}	&	\qw	\\
\ket{0} &	&	\qw	&	\qw	&	\ghost{A}	&	\qw	&		&		&		&	\qw	&	\qw	&	\ghost{A}	&	\qw	\\
	&		&		&		&		&		&		&		&		&		&		&		\\
\vdots	&		&		&	\vdots	&		&		&	\ddots	&		&		&		&		&	\vdots	\\
	&		&		&		&		&		&		&		&		&		&		&		\\
\ket{0} &	&	\gate{X}	&	\qw	&	\multigate{1}{A}	&	\qw	&		&		&		&	\qw	&	\qw	&	\multigate{1}{A}	&	\qw	\\
\ket{0} &	&	\qw	&	\multigate{1}{A}	&	\ghost{A}	&	\qw	&		&		&		&	\qw	&	\multigate{1}{A}	&	\ghost{A}	&	\qw	\\
\ket{0} &	&	\gate{X}	&	\ghost{A}	&	\qw	&	\qw	&		&		&		&	\qw	&	\ghost{A}	&	\qw	&	\qw
} \]
\caption{General construction of an efficient circuit which also enforces number symmetry for any number of qubits (orbitals) $n$ and excitations (electrons) $m$ and is constructed using the logic discussed in the text. Each $A$ gate contributes two variational parameters $\theta_i,\phi_i$ to the ansatz, except the last two $A$ gates, which each contribute one (see text). This general structure only requires single-qubit $X$ gates and a cascade of two-qubit $A$ gates and always generates circuits with the minimal number of required parameters $2\binom{n}{m}-2$. }
\label{fig:nqme}
\end{figure}
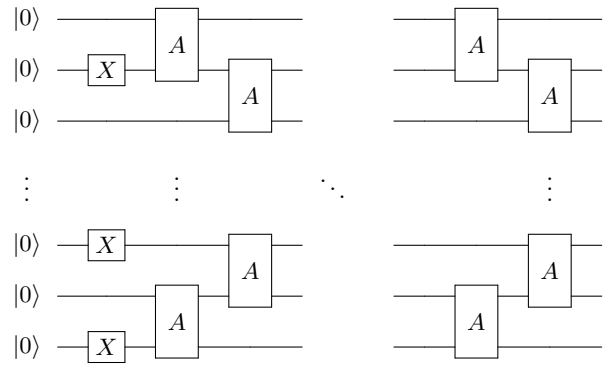

An efficient circuit for the general case of $n$ orbitals and $m$ fermions is shown in Fig.~\ref{fig:nqme}.
This circuit builds on the approach of Fig.~\ref{fig:4q2e}, where $X$ gates are applied to $m$ qubits to bring the system into the correct particle-number subspace, and then a sequence of $A$ gates is performed to create different superpositions. Through trial and error, we find that the total number of $A$ gates needed to produce all possible superpositions is $\binom{n}{m}$. Since each $A$ gate contributes two parameters, we have a total of $2\binom{n}{m}$ parameters; the $\phi$ parameters in the last two $A$ gates can be fixed to reduce the number of parameters down to the minimal number, $2\binom{n}{m}-2$, while still fully spanning the subspace. Our general recipe for constructing circuits for any $n,m$ like that shown in Fig.~\ref{fig:nqme} can be summarized into the following steps:
\begin{enumerate}
        \item{Apply $X$ gates to $m$ qubits. For an efficient circuit, avoid placing $X$ gates on neighboring qubits.}
        \item{Apply a ``first layer" of $A$ gates on all adjacent pairs of qubits on which either $X \otimes \mathds{1}$ or $\mathds{1} \otimes X$ has been applied.}
        \item{Apply a ``second layer" of $A$ gates on adjacent pairs of qubits, where each pair includes one qubit acted on by an $A$ gate from the previous step and a qubit free of $A$ gates. Continue to place $A$ gates on adjacent qubits as necessary until all neighboring qubits are connected with $A$ gates. The first and second layers define a primitive pattern.}
        \item{Repeat the primitive pattern until $\binom{n}{m}$ $A$ gates are placed. Any two $A$ gates have $\phi$ as a free parameter and therefore the full circuit contains exactly $2\binom{n}{m}-2$ parameters, the minimum required to span the subspace $H_{n,m}$.}
\end{enumerate}
In Step 1 above, we can see that this step simply places the full system into the proper $m$ excitation subspace. The following two steps, which only involve applications of $A$ gates, do not change the excitation number of the system but instead mix the $\ket{01}$ and $\ket{10}$ basis states of the two qubits on which they act, producing the desired subspace spanned by $m$, dependent on the parameters $(\vec{\theta},\vec{\phi})$. Although we do not have an analytical argument for why this particular arrangement of $A$ gates works, we have confirmed this through extensive numerical testing. We conjecture that this circuit pattern will continue to work for arbitrarily many fermions and orbitals. Another key point is contained in Step 4, specifically that this general construction always uses the minimal number of required parameters to span the desired space; increasing the gate depth any further is unnecessary, as unit fidelity is achievable at this gate depth. Time-reversal symmetry can be imposed by setting all the $\phi$ parameters and one $\theta$ parameter in the $A$ gates to zero as before. This yields a purely real ansatz with the minimal number of parameters needed. 
Numerical calculations of the fidelity as a function of the number of variational parameters for several different values of $n$ and $m$ are given in Fig. S4 of the supplement~\cite{GardSup2019}. 
We have numerically verified all cases for which ($n>m$) and all permutations of $n=\{2,3,4,5,6\},m=\{1,2,3,4,5\}$.
In all cases, the fidelity increases monotonically with the number of parameters and saturates at unity once this number equals the dimension of the symmetry subspace, confirming that the circuits fully span the subspace.

While this construction is not necessarily the most resource efficient in terms of the required number of gates, it is straightforward to extend to any desired subspace defined by $n$ and $m$ and only requires nearest-neighbor qubit coupling, which is typically more straightforward to engineer. This should be contrasted with the circuit of Fig.~\ref{fig:4q2ev2}, where two-qubit gates between non-neighboring qubits are required. We also note that our general construction naturally exhibits the symmetries of binomial functions. For example, the number of gates used for the case of $\binom{n}{p}$ and $\binom{n}{n-p}$ for $0<p<n$ are identical, which is a reflection of particle-hole symmetry. This in turn means that we can focus on the case where $m\le n/2$ without loss of generality, so that in Step 1 above, we can always avoid applying $X$ gates on two adjacent qubits.

\begin{figure}[!htb]
    \centering
    \begin{varwidth}{9cm}
    \definecolor{mmaBlue}{HTML}{5e81b5}
    \definecolor{mmaOrange}{HTML}{e19c24}
    \definecolor{mmaGreen}{HTML}{8fb032}
    \definecolor{mmaRed}{HTML}{eb6235}
    \definecolor{mmaPurple}{HTML}{8778b3}
    \definecolor{mmaBrown}{HTML}{c56e1a}
    \begin{tikzpicture}
        \begin{axis}[
            xlabel = {$m$},
            ylabel = {\# CNOT},
            xmin = 1,
            xmax = 39,
            xtick = {5, 10, 15, 20, 25, 30, 35},
            ymode=log,
            legend style={at={(axis cs:20,10^4)},anchor=south west}
            ]
            \addplot[line width=0.75pt,solid,color=mmaBlue,mark=*]
                table{
                    1	80
2	2298
3	29596
4	274124
5	1.97E+06
6	1.15E+07
7	5.59E+07
8	2.31E+08
9	8.20E+08
10	2.54E+09
11	6.94E+09
12	1.68E+10
13	3.61E+10
14	6.96E+10
15	1.21E+11
16	1.89E+11
17	2.66E+11
18	3.40E+11
19	3.94E+11
20	4.14E+11
21	3.94E+11
22	3.40E+11
23	2.66E+11
24	1.89E+11
25	1.21E+11
26	6.96E+10
27	3.61E+10
28	1.68E+10
29	6.94E+09
30	2.54E+09
31	8.20E+08
32	2.31E+08
33	5.59E+07
34	1.15E+07
35	1.97E+06
36	274124
37	29596
38	2298
39	80
                };
            \addplot[line width=0.75pt,solid,color=mmaOrange,mark=o]
                table{
                    1	80
2	3200
3	85333.33333
4	1.71E+06
5	2.73E+07
6	3.64E+08
7	4.16E+09
8	4.16E+10
9	3.70E+11
10	2.96E+12
11	2.15E+13
12	1.43E+14
13	8.83E+14
14	5.04E+15
15	2.69E+16
16	1.35E+17
17	6.33E+17
18	2.81E+18
19	1.18E+19
20	4.74E+19
21	1.81E+20
22	6.56E+20
23	2.28E+21
24	7.61E+21
25	2.44E+22
26	7.49E+22
27	2.22E+23
28	6.34E+23
29	1.75E+24
30	4.67E+24
31	1.20E+25
32	3.01E+25
33	7.30E+25
34	1.72E+26
35	3.93E+26
36	8.72E+26
37	1.89E+27
38	3.97E+27
39	8.15E+27
                };        
            \addplot[line width=0.75pt,solid,color=mmaGreen,mark=asterisk]
                table{
                    1	2.20E+12
2	2.20E+12
3	2.20E+12
4	2.20E+12
5	2.20E+12
6	2.20E+12
7	2.20E+12
8	2.20E+12
9	2.20E+12
10	2.20E+12
11	2.20E+12
12	2.20E+12
13	2.20E+12
14	2.20E+12
15	2.20E+12
16	2.20E+12
17	2.20E+12
18	2.20E+12
19	2.20E+12
20	2.20E+12
21	2.20E+12
22	2.20E+12
23	2.20E+12
24	2.20E+12
25	2.20E+12
26	2.20E+12
27	2.20E+12
28	2.20E+12
29	2.20E+12
30	2.20E+12
31	2.20E+12
32	2.20E+12
33	2.20E+12
34	2.20E+12
35	2.20E+12
36	2.20E+12
37	2.20E+12
38	2.20E+12
39	2.20E+12
                };  
            \addplot[line width=0.75pt,solid,color=mmaRed,mark=square]
                table{
                    1	2.56E+06
2	9.73E+08
3	1.56E+11
4	1.34E+13
5	6.93E+14
6	2.36E+16
7	5.56E+17
8	9.46E+18
9	1.20E+20
10	1.15E+21
11	8.55E+21
12	4.99E+22
13	2.32E+23
14	8.62E+23
15	2.59E+24
16	6.32E+24
17	1.26E+25
18	2.06E+25
19	2.76E+25
20	3.04E+25
21	2.76E+25
22	2.06E+25
23	1.26E+25
24	6.32E+24
25	2.59E+24
26	8.62E+23
27	2.32E+23
28	4.99E+22
29	8.55E+21
30	1.15E+21
31	1.20E+20
32	9.46E+18
33	5.56E+17
34	2.36E+16
35	6.93E+14
36	1.34E+13
37	1.56E+11
38	9.73E+08
39	2.56E+06
                };
        \end{axis}
    \end{tikzpicture}
\end{varwidth}
    \caption{Number of required CNOT gates as a function of particle number $m$ for a fixed number of qubits $n=40$. Our general state preparation circuits that respect particle-number symmetry are shown in blue, while the results of Wang {\textit{et. al.}} \cite{Wang2009} are shown in yellow, Bergholm {\textit{et. al.}} \cite{Bergholm2005} in green and Ortiz {\textit{et. al.}} \cite{Ortiz2001} in red.}
    \label{fig:scaling}
\end{figure}
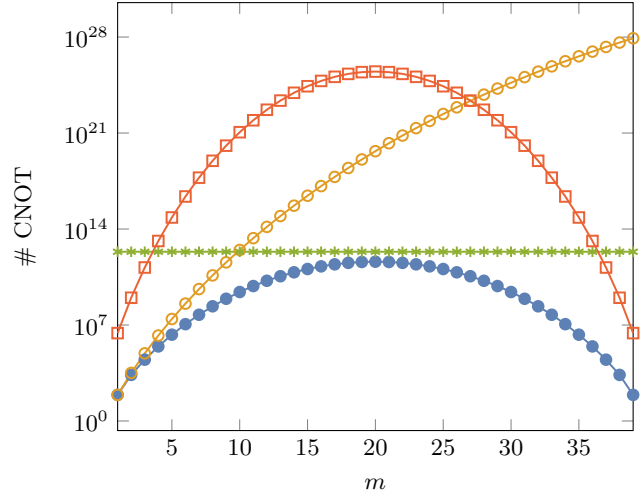

It is worth comparing our general construction with existing state preparation algorithms in terms of gate counts. If we wished to span the full Hilbert space, then this would require $\mathcal{O}(2^n)$ CNOT gates \cite{Shende2005,Mottonen2005}. Some previous state preparation algorithms involve transforming one arbitrary $n-$qubit state into another arbitrary state, which requires $2^{n+1}-2n-2$ CNOT gates \cite{Bergholm2005}. However, since we are interested in only spanning a subspace of the full Hilbert space, namely $H_{n,m}$, we can span this subspace with significantly fewer CNOT gates. Wang {\textit{et. al.}}~\cite{Wang2009} and Ortiz {\textit{et. al.}}~\cite{Ortiz2001} also considered this restricted subspace, requiring no more than $2^{m+1}n^m/m!$ and $\binom{n}{m}^2n^2$ CNOT gates, respectively. In the case of our general construction for $n$ qubits and $m$ excitations, we find that our algorithm requires at most $N_\textrm{CNOT}(n,m)=3\binom{n}{m}$ CNOT gates. Since we always consider a fixed input state (the state with all qubits in $\ket{0}$), simplifications of our required gates are always possible, which reduces the number of required CNOT gates. Specifically, if we eliminate the unnecessary CNOT gates, then the actual number of CNOT gates in the general circuit is 
\begin{equation}
N_{CNOT}(n,m)=
\begin{cases}
  3\binom{n}{m}-3m+1 &0<m<n/2\\
  3\binom{n}{m}-2m-2 &m=n/2\\
  3\binom{n}{m}-3n+3m+1  &n/2<m<n .\\
\end{cases}
\end{equation}
Fig.~\ref{fig:scaling} shows how our approach to constructing state preparation circuits compares to existing works. We see that our scheme significantly decreases the required number of CNOT gates.
We noted earlier that our example of Fig.~\ref{fig:4q2ev2} contains only 9 CNOT gates, while our general construction uses $N_{CNOT}(4,2)=18$ CNOT gates in this case. However,  our general construction in terms of $A$ gates is relatively straightforward for arbitrary $H_{n,m}$, while it is not clear how to generalize the construction of Fig.~\ref{fig:4q2ev2}. We stress that on a case by case basis, it may be possible to find circuits that more efficiently span a reduced Hilbert space, but finding a general procedure for constructing such circuits for arbitrary numbers of orbitals and electrons can be challenging, and it may require a more complicated qubit connectivity beyond just nearest-neighbor coupling. Further investigation of these differences and tradeoffs is an interesting topic for future work but beyond the scope of this paper.

\subsection{Spin Symmetries}\label{sec:spin}

Many systems that we are interested in simulating possess not only particle-number and time-reversal symmetries, but also spin symmetries. This can include both the net spin magnetization $s_z$ and also total spin $s$. To our knowledge, it remains an open problem to find state preparation circuits that respect all these symmetries at the same time. Here, we introduce a protocol for constructing circuits that achieve this.

We begin by first showing how our general scheme for conserving particle number described in the previous section can be extended to conserve $s_z$ as well with only minor modifications. The first step is to choose our fermion-qubit mapping such that the first $n/2$ qubits represent spin up orbitals while the remaining $n/2$ represent spin down orbitals. We follow the same general steps of forming a cascade of $A$ gates as before, but now with the added constraint that the parameters of any $A$ gate that bridges the two spin subspaces (those that entangle the $n/2$ qubit with the $n/2+1$ qubit) are set to zero ($\theta_i=\phi_i=0$). This prevents any mixing of the two spin subspaces. Therefore, if we start with the proper number of spin-up and spin-down orbitals occupied, then this sequence of gates guarantees that the final state also has the correct spin occupation numbers. We also require that these parameter-free $A$ gates do not appear in the first layer. This can effectively swap the first and second layers as outlined in the previous general protocol. For example, in the case of Fig.~\ref{fig:4q2e} and with our spin assignment, we only need to set the four parameters contained within the $A$ gates that act between qubits $2,3$ to zero along with $\phi_5=\phi_4$. This assignment then generates all states with $n=4,m=2,s_z=0$ and with the minimal number of parameters ($6$) for this space.

\begin{figure}[!tb]
\centering
\begin{varwidth}{9cm}
    \definecolor{mmaBlue}{HTML}{5e81b5}
    \definecolor{mmaOrange}{HTML}{e19c24}
    \definecolor{mmaGreen}{HTML}{8fb032}
    \definecolor{mmaRed}{HTML}{eb6235}
    \definecolor{mmaPurple}{HTML}{8778b3}
    \definecolor{mmaBrown}{HTML}{c56e1a}
    \begin{tikzpicture}
        \begin{axis}[
            xlabel = {$n$},
            ylabel = {$\text{dim}(H)$},
            xmin=3,
            xmax=29,
            ymode=log,
            xtick = {4, 8, 12, 16, 20, 24, 28},
            legend style={at={(axis cs:20,10^4)},anchor=south west}
            ]
            \addplot[line width=0.75pt,solid,color=mmaBlue,mark=*]
                table{
                    4	16
                    5	32
                    6	64
                    7	128
                    8	256
                    9	512
                    10	1024
                    11	2048
                    12	4096
                    13	8192
                    14	16384
                    15	32768
                    16	65536
                    17	131072
                    18	262144
                    19	524288
                    20	1048576
                    21	2097152
                    22	4194304
                    23	8388608
                    24	16777216
                    25	33554432
                    26	67108864
                    27	134217728
                    28	268435456
                };           
            \addplot[line width=0.75pt,solid,color=mmaOrange,mark=o]
                table{
                    4	6
                    5	10
                    6	20
                    7	35
                    8	70
                    9	126
                    10	252
                    11	462
                    12	924
                    13	1716
                    14	3432
                    15	6435
                    16	12870
                    17	24310
                    18	48620
                    19	92378
                    20	184756
                    21	352716
                    22	705432
                    23	1352078
                    24	2704156
                    25	5200300
                    26	10400600
                    27	20058300
                    28	40116600
                };           
            \addplot[line width=0.75pt,solid,color=mmaGreen,mark=asterisk]
                table{
                    4    3
                    8    20
                    12   175
                    16   1764
                    20   19404
                    24   226512
                    28   2760615
                };         
            \addplot[line width=0.75pt,solid,color=mmaRed,mark=square]
                table{
                    4    4
                    8    36
                    12   400
                    16   4900
                    20   63504
                    24   853776
                    28   11778624
                };          
        \end{axis}
    \end{tikzpicture}
\end{varwidth}
\caption{(Color Online) Hilbert space dimension as a function of number of qubits, $n$, when relevant symmetries are enforced. We show the dimension of the full Hilbert space (blue), the largest particle-number subspace with $m=n/2$ (yellow), the subspace with $m=n/2$ and $s_z=0$ (red), and the subspace with $m=n/2$ and $s=0$ (green). Lines are included only as a guide. All cases remain an exponential scaling Hilbert space, where use of symmetries reduces the exponential factor.}
\label{fig:dim}
\end{figure}
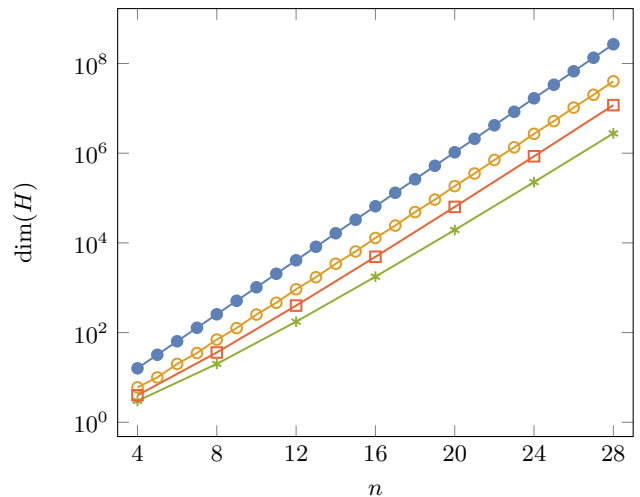

We now move on to the case where total spin $s$ is also a good quantum number.
If we impose conservation of $s$, $s_z$ and particle number, then the size of the relevant symmetry subspace for a chosen $n,m,s,s_z$ is given by
\begin{equation}
\begin{split}
\textrm{dim}(H_{n,m,s,s_z})&=
\sum_{k=0}^{m/2-s} \binom{n/2}{k} \binom{n/2-k}{m-2k}\\
&\times\frac{(2s+1)(m-2k)!}{(m/2-k-s)!(m/2-k+s+1)!}.
 \label{dimspin}
 \end{split}
\end{equation}
Here, we maintain the notation from the previous section where $n$ is the number of qubits and $m$ is the number of particles, but now we focus on the case in which each qubit encodes the occupancy of a particular spin-orbital. (In the previous section each qubit could correspond to either a spin-orbital or a spatial orbital in the case of spinless fermions.) Therefore, the $n$ qubits encode $n/2$ spatial orbitals, and we omit the case of odd values for $n$. The first binomial coefficient in Eq.~\eqref{dimspin} counts the number of ways to assign doubly occupied spatial orbitals in the system, the second counts the number of ways to assign singly occupied orbitals, and the last factor is the number of $s$ irreducible representations in a tensor product of $m$ spin-$\frac{1}{2}$ particles. We show in Fig.~\ref{fig:dim} that exploiting all the symmetries may significantly reduce the dimension of the Hilbert space. For example, already for $n=28$ spin-orbitals, the relevant subspace when all symmetries are imposed is at least two orders of magnitude smaller than the full Hilbert space of $28$ qubits. Taking advantage of this reduction can significantly reduce the demands on the quantum processor and improve the speed and accuracy of the classical optimization step of the VQE algorithm. 
We note that, while this reduction is significant, this scaling always remains exponential even when symmetries are imposed. Nevertheless, use of symmetries can simplify ans\"atze and thus reduce the required CNOT and parameter counts.
The problem of constructing circuits that also conserve total spin $s$ is much more challenging compared to the symmetries we have discussed thus far. In the remainder of this section, we address this problem by developing a completely different approach to obtaining symmetry-preserving circuits. In addition to allowing for conservation of $s$, this approach also provides an alternative method for building circuits that respect particle-number symmetry. In this new approach, we adopt the convention that the qubits alternate between spin-up and spin-down orbitals such that the first, third, fifth, etc. qubits encode spin up, while the second, fourth, sixth, etc. encode spin down. This choice of spin encoding leads to an efficient scaling in terms of number of gates, as we will see. 

The starting point for this alternative circuit-building scheme is to find an $n$-qubit unitary that transforms the state $\otimes_{i=1}^n\ket{0}$ into an arbitrary superposition of states that share the same value of $m$. This is of course exactly the problem of finding a circuit that respects particle-number symmetry that we discussed extensively in the previous section. The key idea here is that by starting with the explicit $n$-qubit unitary upfront instead of a decomposition of it in terms of $A$ gates, we can control how circuit parameters appear in the coefficients of the final superposition states. If the circuit parameters appear in a sufficiently simple way, then it would be straightforward to impose constraints on them so that spin symmetries are also respected. We could attempt to use the particle-number circuits from the previous section and impose constraints on the circuit parameters to enforce total spin symmetry, but this would lead to complicated, highly non-linear conditions on the parameters that would be difficult to solve.

It is straightforward to construct an $n$-qubit unitary that transforms $\otimes_{i=1}^n\ket{0}$ into an arbitrary state in a particular particle-number subspace. In the simplest case of only two spin-orbitals and a single fermion, this unitary is just the $A$ gate multiplied by a single $X$ gate (see Fig.~\ref{fig:2q1e}). In the case of four spin-orbitals and two fermions, the desired unitary should generate arbitrary superposition states formed from a total of $\binom{4}{2}=6$ basis states. While many different unitaries transform $\otimes_{i=1}^n\ket{0}$ into such superposition states, here we desire a unitary which generates these states with {\it{specific}} coefficients that are easily adjusted to respect the appropriate spin symmetries. The unitary we choose to use, referred to as the $E_4$ gate, is based on hyperspherical coordinates:
\begin{equation}
\begin{split}
E_4\ket{0000}=&\sin{u_1} \sin{u_2} \sin{u_3} \sin{u_4} \sin{u_5}\ket{0101}\\
+&\sin{u_1} \sin{u_2} \sin{u_3} \sin{u_4} \cos{u_5}\ket{1001}\\
+&\sin{u_1} \sin{u_2} \sin{u_3} \cos{u_4}\ket{0011}\\
+&\sin{u_1}\sin{u_2}\cos{u_3}\ket{0110}\\
+&\sin{u_1}\cos{u_2}\ket{1010}\\
+&\cos{u_1}\ket{1100} .
\end{split}
\label{eq:egateaction}
\end{equation}
The $E_4$ gate clearly generates any two-particle state in a system with four spin-orbitals. Notice that we have also imposed time-reversal symmetry by purposely choosing the coefficients to be real. This can be easily generalized to problems without time-reversal symmetry by inserting additional arbitrary phase factors on any five of the six terms in Eq.~\eqref{eq:egateaction}. The use of hyperspherical coordinates provides a simple parameterization that automatically ensures normalization and facilitates the incorporation of spin symmetries. For example, if we want to restrict to the subspace with $s=1,s_z=0$, then there is only one state:
\[\ket{s=1,s_z=0}=\frac{1}{\sqrt{2}}(\ket{1001}+\ket{0110}).
\]
This state can be created from the $E_4$ gate by setting $u_1=u_2=u_4=\pi/2,u_3=\pi/4,u_5=0$. We summarize this case and the other spin subspace of interest $s=0,s_z=0$, which is given by the general superposition
\begin{equation}\ket{s=0,s_z=0}=\frac{\gamma}{\sqrt{2}}(\ket{1001}-\ket{0110})+\alpha \ket{1100}+\beta \ket{0011},
\label{eq:s0sz0}
\end{equation}
in Table~\ref{tbl1}.

The advantage of this construction is that we can easily fix parameters in the $E_4$ gate to generate any desired spin subspace in this $n=4,m=2$ space. There are of course many unitaries that satisfy Eq.~\eqref{eq:egateaction}. Once we have settled on a particular choice for $E_4$, we can then decompose it into more elementary gates. Note that in Eq.~\eqref{eq:egateaction} we have chosen a basis ordering which leads to an efficient Gray code decomposition~\cite{Nielsen2011} (there are only two bit changes from term to term). A specific choice for $E_4$ and its corresponding decomposition are shown in the appendix.

The case of $n>4$ spin-orbitals can be treated in a similar fashion. We begin by constructing a unitary $E_n$ that transforms $\otimes_{i=1}^n\ket{0}$ into an arbitrary superposition of $\binom{n}{m}$ fixed-particle-number basis states. For example, in the case $n=6,m=3$, we have
\begin{equation}
    \begin{split}E_6\ket{000000}=&
        \sum_{j=20}^{1}\cos{u_j}\prod_{i=1}^{j-1}\sin{u_i} \ket{p_{j1},p_{j2},...,p_{j6}},
    \end{split}
    \label{eq:fgate}
\end{equation}
where $p_{jk}\in\{0,1\}$, $\sum_{k=1}^6 p_{jk}=3$, $u_{20}=0$, and the order of the basis states is given by
\begin{equation}
    \begin{split}
&\ket{000111},\ket{001011},\ket{001110},\ket{001101},\ket{011001},\ket{011010}\\
&\ket{011100},\ket{010110},\ket{010101},\ket{010011},\ket{100011},\ket{100101}\\
&\ket{100110},\ket{101010},\ket{101100},\ket{101001},\ket{110001},\ket{110010}\\
&\ket{110100},\ket{111000}.
    \end{split}
\end{equation}
This ordering results in minimal (2) bit changes per step of the Gray code, which facilitates the decomposition of $E_6$. We list the specific spin subspaces and summarize how to generate them by fixing the parameters of $E_6$ in Table.~\ref{tbl2}.

An example of a gate decomposition for a specific choice of $E_6$ is given in the appendix along with explicit gate counts for $E_4$, $E_6$, and $E_8$.

Extrapolating from these examples, we can then form a general procedure for constructing symmetry-preserving state preparation circuits for any valid choice of the quantum numbers $n,m,s,s_z$ for a time-reversal-symmetric system as follows:
\begin{enumerate}
    \item {For a given choice of $n,m$, there are $\binom{n}{m}$ basis states ($\ket{p_{j1},p_{j2},...,p_{jn}}$) that span the corresponding particle-number subspace, such that $\sum_{k=1}^n p_{jk}=m$. Assign hyperspherical coefficients to these basis states according to $\sum_{j=\binom{n}{m}}^{1}\cos{u_j}\prod_{i=1}^{j-1}\sin{u_i} \ket{p_{j1},p_{j2},...,p_{jn}}$, where $u_{\binom{n}{m}}=0$, and the basis states are ordered such that there are only two bit changes per step in the Gray code.}
    \item{Determine the constraints that must be imposed on the $u_i$ such that the appropriate spin subspace labled by $s$ and $s_z$ is obtained.}
    \item{Construct a unitary $E_n$ such that the first column contains the coefficients in Step 1. The remaining columns can be chosen as desired so long as they respect unitarity.}
    \item{Decompose $E_n$ using a Gray code scheme or alternative gate decomposition technique.}
\end{enumerate}
Following this procedure, we can enforce arbitrary spin and particle-number symmetries for any choice of $n,m,s,s_z$. Step 2 is facilitated by the use of hyperspherical coordinates because various basis states are easily eliminated from the final superposition by setting the corresponding $u_j$ to $\pi/2$, yielding another hypersphere parameterization of lower dimension. This recursive structure in the coefficients maintains regularity and simplicity as the effective Hilbert space dimension is reduced. Note that in cases where it is necessary to fix some of the $u_j$ in terms of the others (see e.g., the case of $s=s_z=0$ in Table~\ref{tbl1} or $s=1/2=\pm s_z$ in Table~\ref{tbl2}), use of inverse functions without a restricted real-domain (e.g. $\tan^{-1},\cot^{-1}$) are desirable for optimization stability and enforcement of time-reversal symmetry, while the use of other inverse trigonometric functions could yield complex-valued coefficients in the trial states. Finally, we emphasize that here it is not necessary to numerically compute fidelities to confirm that the resulting circuits indeed span the appropriate symmetry subspace as in the previous section because this is guaranteed by construction in the present approach. 
\begin{figure*}[!ht]
    \definecolor{mmaBlue}{HTML}{5e81b5}
    \definecolor{mmaOrange}{HTML}{e19c24}
    \definecolor{mmaGreen}{HTML}{8fb032}
    \definecolor{mmaRed}{HTML}{eb6235}
    \definecolor{mmaPurple}{HTML}{8778b3}
    \definecolor{mmaBrown}{HTML}{c56e1a}
    \definecolor{mmaBlack}{HTML}{696969}
    \centering
    \begin{tikzpicture}[baseline=2.8 cm]
        \begin{axis}[
            xlabel = { Interatomic Distance [\AA]},
            ylabel = {$\Delta$E (Hartree)},
            xmin = 0,
            xmax = 4.2,
            ymin= 1e-16,
            ymax=1e2,
            ymode=log,
            legend style={at={(0,1)},anchor=north west, 
            legend columns =1}
            ]
        \addlegendentry{$A_{4,2,s_z=0}$}
        \addlegendentry{$E_{4,2}$}
        \addlegendentry{$E_{4,2,s=0,s_z=0}$}
        \addlegendentry{UCCSD}
        \addlegendentry{SWAPRZ}
        \addlegendentry{RYRZ}
        \addlegendentry{RY}
             \addplot[line width=0.75pt,solid,color=mmaBlue,mark=*]
                table{
                  0.2 2.4913404672588513e-13
0.25507246376811593 1.0009770790020411e-12
0.3101449275362319 2.1316282072803006e-13
0.3652173913043478 6.008527009271347e-13
0.42028985507246375 1.332711718760038e-12
0.4753623188405797 5.284661597215745e-14
0.5304347826086957 8.526512829121202e-14
0.5855072463768116 2.5757174171303632e-14
0.6405797101449275 8.659739592076221e-15
0.6956521739130435 3.1086244689504383e-15
0.7507246376811594 3.2445157671645575e-12
0.8057971014492753 8.881784197001252e-16
0.8608695652173912 3.779199175824033e-13
0.9159420289855071 2.8532731732866523e-13
0.9710144927536233 2.717825964282383e-13
1.0260869565217392 8.65529869997772e-13
1.0811594202898551 6.994405055138486e-14
1.136231884057971 6.991740519879386e-12
1.191304347826087 2.0479173912235638e-12
1.2463768115942029 1.1146639167236572e-13
1.3014492753623188 2.8883562208648073e-12
1.3565217391304347 7.105427357601002e-15
1.4115942028985506 5.88418203051333e-13
1.4666666666666666 4.1677772344428377e-13
1.5217391304347825 3.5438318946034997e-13
1.5768115942028984 4.547473508864641e-12
1.6318840579710143 6.661338147750939e-16
1.6869565217391305 1.2723155862204294e-13
1.7420289855072464 2.191580250610059e-13
1.7971014492753623 2.4868995751603507e-14
1.8521739130434782 1.4832579608992091e-12
1.9072463768115941 1.1546319456101628e-14
1.96231884057971 8.881784197001252e-16
2.017391304347826 1.6120438317557273e-13
2.072463768115942 4.656275365277907e-13
2.127536231884058 5.995204332975845e-14
2.1826086956521737 2.4003021792395884e-13
2.23768115942029 1.5398793351550921e-12
2.292753623188406 1.908251334725719e-12
2.347826086956522 1.509903313490213e-13
2.402898550724638 1.822986206434507e-13
2.457971014492754 1.638689184346731e-13
2.5130434782608697 1.7899015603006774e-12
2.5681159420289856 5.351274978693255e-13
2.6231884057971016 2.930988785010413e-14
2.6782608695652175 2.9531932455029164e-14
2.7333333333333334 5.412337245047638e-12
2.7884057971014493 4.340972026284362e-13
2.8434782608695652 1.9984014443252818e-15
2.898550724637681 7.30304705598428e-13
2.953623188405797 1.028066520802895e-13
3.008695652173913 2.970956813896919e-13
3.063768115942029 1.1884826456309838e-10
3.118840579710145 1.0196732347367288e-11
3.173913043478261 7.402967128200544e-13
3.228985507246377 9.57900425646585e-13
3.284057971014493 8.599787548746463e-13
3.339130434782609 1.0935030658743017e-11
3.394202898550725 3.235189893757706e-13
3.449275362318841 5.659916979539048e-13
3.5043478260869567 8.586464872450961e-13
3.5594202898550726 1.0620837542774098e-11
3.6144927536231886 1.744826505500896e-12
3.6695652173913045 4.298783551348606e-13
3.7246376811594204 1.2049694575466674e-11
3.7797101449275363 2.7500224319965128e-12
3.8347826086956522 2.2483570560893895e-11
3.889855072463768 1.5552004128949193e-11
3.944927536231884 9.024114788758197e-12
4.0 2.969668955188354e-11
                   };
                    \addplot[line width=0.75pt,solid,color=mmaGreen,mark=+]
                table{
                  0.2 2.6645352591003757e-15
0.25507246376811593 8.304468224196171e-14
0.3101449275362319 1.4788170688007085e-13
0.3652173913043478 4.440892098500626e-14
0.42028985507246375 7.194245199571014e-14
0.4753623188405797 1.7763568394002505e-14
0.5304347826086957 2.708944180085382e-14
0.5855072463768116 2.589040093425865e-13
0.6405797101449275 1.7763568394002505e-14
0.6956521739130435 2.7000623958883807e-13
0.7507246376811594 4.483080573436382e-13
0.8057971014492753 3.3306690738754696e-14
0.8608695652173912 4.440892098500626e-15
0.9159420289855071 5.750955267558311e-14
0.9710144927536233 4.6629367034256575e-14
1.0260869565217392 1.070254995738651e-12
1.0811594202898551 3.3240077357277187e-13
1.136231884057971 3.1885605267234496e-13
1.191304347826087 6.770140004164205e-13
1.2463768115942029 1.071320809842291e-11
1.3014492753623188 6.048495038157853e-12
1.3565217391304347 2.2073454175597362e-12
1.4115942028985506 3.375077994860476e-14
1.4666666666666666 3.028688411177427e-12
1.5217391304347825 2.0228263508670352e-13
1.5768115942028984 3.930189507173054e-14
1.6318840579710143 2.0805579481475434e-13
1.6869565217391305 2.7817748105007922e-12
1.7420289855072464 3.9745984281580604e-14
1.7971014492753623 9.805711798094308e-12
1.8521739130434782 7.054357098468245e-12
1.9072463768115941 4.807487741231853e-12
1.96231884057971 2.4069635173873394e-13
2.017391304347826 3.3657521214536246e-12
2.072463768115942 5.271338920920243e-12
2.127536231884058 1.3131717935266352e-12
2.1826086956521737 6.816769371198461e-13
2.23768115942029 5.0459636469213365e-12
2.292753623188406 3.326583453144849e-11
2.347826086956522 6.364020421756322e-12
2.402898550724638 3.630651335129187e-12
2.457971014492754 9.15045816896054e-13
2.5130434782608697 5.597966534764964e-12
2.5681159420289856 2.8328450696335494e-11
2.6231884057971016 2.1476820322163803e-11
2.6782608695652175 2.220446049250313e-14
2.7333333333333334 1.192379528447418e-13
2.7884057971014493 2.883226990491039e-11
2.8434782608695652 5.263478541905897e-11
2.898550724637681 5.635159006089907e-11
2.953623188405797 5.5104587559640095e-11
3.008695652173913 2.489564110419451e-12
3.063768115942029 1.27675647831893e-11
3.118840579710145 1.8419710201555972e-11
3.173913043478261 2.185807090882008e-12
3.228985507246377 1.2962231288327075e-10
3.284057971014493 4.3926817738793034e-10
3.339130434782609 2.427213985356502e-10
3.394202898550725 3.3626212925241816e-11
3.449275362318841 1.6200973895763582e-10
3.5043478260869567 1.8928958400721285e-09
3.5594202898550726 1.1427525592466736e-11
3.6144927536231886 5.671862979284015e-11
3.6695652173913045 1.013882311440284e-10
3.7246376811594204 4.091189609312096e-10
3.7797101449275363 3.1185229953933913e-09
3.8347826086956522 4.928019547989493e-09
3.889855072463768 1.5459878044410402e-08
3.944927536231884 7.165021242983016e-09
4.0 3.3119937159487023e-06
                   };
                      \addplot[line width=0.75pt,solid,color=mmaRed,mark=x]
                table{
                  0.2 8.43769498715119e-15
0.25507246376811593 4.440892098500626e-15
0.3101449275362319 7.105427357601002e-15
0.3652173913043478 4.884981308350689e-15
0.42028985507246375 1.865174681370263e-14
0.4753623188405797 1.7763568394002505e-15
0.5304347826086957 5.773159728050814e-15
0.5855072463768116 7.105427357601002e-15
0.6405797101449275 8.881784197001252e-16
0.6956521739130435 2.6645352591003757e-15
0.7507246376811594 1.0247358517290195e-12
0.8057971014492753 1.3322676295501878e-15
0.8608695652173912 1.2212453270876722e-14
0.9159420289855071 7.327471962526033e-15
0.9710144927536233 1.3308465440786676e-11
1.0260869565217392 6.039613253960852e-14
1.0811594202898551 7.815970093361102e-14
1.136231884057971 1.3322676295501878e-15
1.191304347826087 1.687538997430238e-14
1.2463768115942029 7.993605777301127e-15
1.3014492753623188 8.881784197001252e-16
1.3565217391304347 3.9968028886505635e-15
1.4115942028985506 1.7763568394002505e-14
1.4666666666666666 1.1102230246251565e-14
1.5217391304347825 1.7763568394002505e-15
1.5768115942028984 1.509903313490213e-14
1.6318840579710143 7.771561172376096e-15
1.6869565217391305 9.992007221626409e-15
1.7420289855072464 3.9479530755670567e-13
1.7971014492753623 4.884981308350689e-15
1.8521739130434782 4.440892098500626e-16
1.9072463768115941 1.1546319456101628e-14
1.96231884057971 8.43769498715119e-15
2.017391304347826 5.329070518200751e-15
2.072463768115942 4.218847493575595e-15
2.127536231884058 9.92850246461785e-12
2.1826086956521737 2.4424906541753444e-15
2.23768115942029 4.4853010194856324e-14
2.292753623188406 2.1760371282653068e-14
2.347826086956522 2.318145675417327e-13
2.402898550724638 2.0872192862952943e-14
2.457971014492754 5.773159728050814e-15
2.5130434782608697 1.1324274851176597e-14
2.5681159420289856 1.9984014443252818e-15
2.6231884057971016 1.9095836023552692e-14
2.6782608695652175 1.709743457922741e-14
2.7333333333333334 5.764277943853813e-13
2.7884057971014493 3.552713678800501e-15
2.8434782608695652 3.419486915845482e-14
2.898550724637681 5.329070518200751e-15
2.953623188405797 1.5543122344752192e-15
3.008695652173913 1.199040866595169e-14
3.063768115942029 2.1760371282653068e-14
3.118840579710145 1.865174681370263e-14
3.173913043478261 9.50350909079134e-14
3.228985507246377 2.220446049250313e-16
3.284057971014493 5.242251077675064e-12
3.339130434782609 1.4876988529977098e-14
3.394202898550725 9.325873406851315e-15
3.449275362318841 1.3988810110276972e-14
3.5043478260869567 1.9984014443252818e-15
3.5594202898550726 4.6629367034256575e-15
3.6144927536231886 6.306066779870889e-14
3.6695652173913045 2.220446049250313e-15
3.7246376811594204 8.215650382226158e-15
3.7797101449275363 1.3322676295501878e-15
3.8347826086956522 9.103828801926284e-14
3.889855072463768 6.661338147750939e-16
3.944927536231884 7.993605777301127e-15
4.0 7.771561172376096e-15
                   };
                   \addplot[line width=0.75pt,solid,color=mmaOrange,mark=square]
                table{
                  0.2 4.206412995699793e-12
0.25507246376811593 6.156408716151418e-12
0.3101449275362319 2.4424906541753444e-14
0.3652173913043478 1.509903313490213e-14
0.42028985507246375 1.9539925233402755e-14
0.4753623188405797 1.3322676295501878e-14
0.5304347826086957 8.881784197001252e-15
0.5855072463768116 7.105427357601002e-15
0.6405797101449275 1.8207657603852567e-14
0.6956521739130435 1.9984014443252818e-14
0.7507246376811594 7.993605777301127e-15
0.8057971014492753 1.0658141036401503e-14
0.8608695652173912 1.0436096431476471e-14
0.9159420289855071 6.439293542825908e-15
0.9710144927536233 2.6645352591003757e-15
1.0260869565217392 4.440892098500626e-16
1.0811594202898551 7.753797603982093e-13
1.136231884057971 1.509903313490213e-14
1.191304347826087 9.992007221626409e-15
1.2463768115942029 7.993605777301127e-15
1.3014492753623188 5.329070518200751e-15
1.3565217391304347 0.0
1.4115942028985506 5.773159728050814e-15
1.4666666666666666 1.2656542480726785e-14
1.5217391304347825 1.021405182655144e-14
1.5768115942028984 1.509903313490213e-14
1.6318840579710143 9.992007221626409e-15
1.6869565217391305 9.970468894948681e-12
1.7420289855072464 3.5482727867020003e-13
1.7971014492753623 1.199040866595169e-14
1.8521739130434782 2.220446049250313e-16
1.9072463768115941 2.219113781620763e-12
1.96231884057971 3.572142581731441e-11
2.017391304347826 2.886579864025407e-15
2.072463768115942 1.0658141036401503e-14
2.127536231884058 9.325873406851315e-15
2.1826086956521737 8.215650382226158e-15
2.23768115942029 1.7763568394002505e-14
2.292753623188406 1.9984014443252818e-15
2.347826086956522 5.950795411990839e-14
2.402898550724638 1.5276668818842154e-13
2.457971014492754 3.5060843117662444e-13
2.5130434782608697 7.294165271787278e-13
2.5681159420289856 9.263700917472306e-13
2.6231884057971016 1.489253165232185e-12
2.6782608695652175 2.418509836843441e-12
2.7333333333333334 2.8170799026838722e-12
2.7884057971014493 3.804068171575636e-12
2.8434782608695652 4.562794586604468e-12
2.898550724637681 5.345945908175054e-12
2.953623188405797 6.066258606551855e-12
3.008695652173913 7.014167024976814e-12
3.063768115942029 7.764011655808645e-12
3.118840579710145 8.002487561498128e-12
3.173913043478261 7.608136343151273e-12
3.228985507246377 8.255174321902814e-12
3.284057971014493 8.031131315533457e-12
3.339130434782609 8.273826068716517e-12
3.394202898550725 7.344347352500336e-12
3.449275362318841 7.297273896256229e-12
3.5043478260869567 7.149614233981083e-12
3.5594202898550726 6.853628775616016e-12
3.6144927536231886 6.489475623538965e-12
3.6695652173913045 6.211031688962976e-12
3.7246376811594204 5.3372861685829776e-12
3.7797101449275363 5.004663350405281e-12
3.8347826086956522 4.7699622029995226e-12
3.889855072463768 4.581002244208321e-12
3.944927536231884 4.143352327901084e-12
4.0 3.782307800292983e-12
                   };
                   \addplot[line width=0.75pt,solid,color=mmaPurple,mark=o]
                table{
                  0.2 7.793765632868599e-13
0.25507246376811593 8.893774605667204e-12
0.3101449275362319 2.8812507935072063e-12
0.3652173913043478 5.1176840543121216e-12
0.42028985507246375 2.2923885012460232e-12
0.4753623188405797 3.632649736573512e-12
0.5304347826086957 7.212008767965017e-13
0.5855072463768116 9.907630271754897e-13
0.6405797101449275 1.5052403767867872e-12
0.6956521739130435 2.361222328772783e-12
0.7507246376811594 6.589617740360154e-12
0.8057971014492753 9.667822098435863e-13
0.8608695652173912 2.3450130726132556e-12
0.9159420289855071 9.18309872588452e-12
0.9710144927536233 6.366907001620348e-12
1.0260869565217392 4.119415919490166e-11
1.0811594202898551 2.8331115231594595e-11
1.136231884057971 2.524469522313666e-11
1.191304347826087 4.148237309209435e-12
1.2463768115942029 6.8620664706031675e-12
1.3014492753623188 1.2527845427712236e-10
1.3565217391304347 5.725109275545037e-11
1.4115942028985506 4.710609680103062e-11
1.4666666666666666 2.5143798154658725e-10
1.5217391304347825 7.121858658365454e-12
1.5768115942028984 3.785838309511291e-11
1.6318840579710143 3.456213093500082e-11
1.6869565217391305 2.0751511620176188e-10
1.7420289855072464 2.19868567796766e-11
1.7971014492753623 1.1590439719100232e-10
1.8521739130434782 1.5093570837620973e-10
1.9072463768115941 9.138045875545231e-11
1.96231884057971 6.26341201126479e-11
2.017391304347826 3.1189784088780925e-10
2.072463768115942 9.337361994710136e-10
2.127536231884058 2.670914600599872e-10
2.1826086956521737 6.751419423522975e-10
2.23768115942029 2.0990431615075522e-10
2.292753623188406 9.394636180104499e-10
2.347826086956522 2.187378056461853e-09
2.402898550724638 9.466714079309213e-10
2.457971014492754 6.718374745418032e-10
2.5130434782608697 8.911715809745147e-10
2.5681159420289856 1.013721995235528e-09
2.6231884057971016 4.2227106256120805e-09
2.6782608695652175 2.9757356578841154e-09
2.7333333333333334 2.4090811567845094e-09
2.7884057971014493 1.2226192724895668e-08
2.8434782608695652 7.67675722990191e-09
2.898550724637681 2.0183744009472093e-08
2.953623188405797 1.0911129333379677e-08
3.008695652173913 4.546098608670945e-09
3.063768115942029 1.5506126782582896e-07
3.118840579710145 2.1142446460231668e-08
3.173913043478261 3.339497989252038e-08
3.228985507246377 1.075519162618832e-08
3.284057971014493 1.1348368600394565e-07
3.339130434782609 3.668774772869199e-08
3.394202898550725 2.272201093944659e-09
3.449275362318841 1.3274819687936201e-08
3.5043478260869567 8.564655429310619e-08
3.5594202898550726 1.2909737723276749e-08
3.6144927536231886 6.723072210057524e-08
3.6695652173913045 5.051420928214867e-07
3.7246376811594204 2.2109686814708596e-07
3.7797101449275363 1.2909165483243612e-06
3.8347826086956522 1.6643779208180831e-06
3.889855072463768 1.1139295339024358e-05
3.944927536231884 5.230831634950306e-06
4.0 2.8483796097322767e-06
                   };
                    \addplot[line width=0.75pt,solid,color=mmaBrown,mark=diamond]
                table{
                  0.2 6.645128891591412e-10
0.25507246376811593 6.375604488795261e-09
0.3101449275362319 5.020655002851981e-10
0.3652173913043478 1.0631007185679664e-10
0.42028985507246375 6.9921504142200774e-09
0.4753623188405797 4.227223016073367e-10
0.5304347826086957 8.265765849557738e-10
0.5855072463768116 8.911196225369622e-10
0.6405797101449275 5.281186599148668e-10
0.6956521739130435 1.3258689701700632e-09
0.7507246376811594 2.079371785868034e-09
0.8057971014492753 1.0330318822582285e-09
0.8608695652173912 2.826137102118764e-10
0.9159420289855071 1.0518563797745628e-10
0.9710144927536233 1.2889214140443528e-09
1.0260869565217392 8.725011824139983e-10
1.0811594202898551 1.6100796251805605e-09
1.136231884057971 3.5331337855382117e-10
1.191304347826087 3.771797985052672e-09
1.2463768115942029 2.532587473069725e-10
1.3014492753623188 4.828226707331851e-10
1.3565217391304347 2.259071374410837e-09
1.4115942028985506 1.0548228956963612e-11
1.4666666666666666 6.450473488683883e-10
1.5217391304347825 3.819471405819286e-10
1.5768115942028984 4.2052361592936904e-11
1.6318840579710143 1.6049699347320256e-09
1.6869565217391305 3.051114916274855e-10
1.7420289855072464 4.1948933215962825e-10
1.7971014492753623 1.1304437386172594e-09
1.8521739130434782 8.389768879624171e-10
1.9072463768115941 2.75183653641875e-10
1.96231884057971 3.115323488067645e-08
2.017391304347826 1.077520295211798e-09
2.072463768115942 7.944491731137759e-10
2.127536231884058 2.102153118244132e-09
2.1826086956521737 2.6357376281538336e-09
2.23768115942029 8.359770431454194e-09
2.292753623188406 1.271049265838542e-09
2.347826086956522 0.0014311153706334512
2.402898550724638 4.35320934855099e-09
2.457971014492754 4.766075090145705e-09
2.5130434782608697 1.007634442551364e-08
2.5681159420289856 0.0017199879946714347
2.6231884057971016 2.7163051807832517e-09
2.6782608695652175 8.619010638533098e-08
2.7333333333333334 0.0019014216061292721
2.7884057971014493 0.0007723569313369261
2.8434782608695652 3.965612762613091e-07
2.898550724637681 1.686471162720693e-08
2.953623188405797 0.000415569251415171
3.008695652173913 6.195781554474422e-08
3.063768115942029 4.6637835793283955e-07
3.118840579710145 2.7185020678999194e-08
3.173913043478261 1.1821132961387093e-08
3.228985507246377 0.00014051546874593868
3.284057971014493 4.113329917121433e-06
3.339130434782609 9.017178684311844e-05
3.394202898550725 7.201338274431812e-05
3.449275362318841 2.3771342472223012e-07
3.5043478260869567 1.6819300239223622e-06
3.5594202898550726 1.0818261790834782e-06
3.6144927536231886 4.233064163794431e-05
3.6695652173913045 2.0294612457227856e-07
3.7246376811594204 3.411254330254465e-05
3.7797101449275363 1.3462256511331816e-05
3.8347826086956522 1.0987308403320739e-05
3.889855072463768 1.074041726667474e-06
3.944927536231884 9.096689371101974e-06
4.0 9.34546384345758e-06
                   };
                   \addplot[line width=0.75pt,solid,color=mmaBlack,mark=pentagon]
                table{
                  0.2 2.90842905315003e-11
0.25507246376811593 1.2397194382174348e-11
0.3101449275362319 1.3876455540184907e-11
0.3652173913043478 5.282432269382298e-10
0.42028985507246375 8.377565308137491e-11
0.4753623188405797 2.1617996281975138e-10
0.5304347826086957 1.294133689100363e-10
0.5855072463768116 1.0254952442778631e-10
0.6405797101449275 2.95033997232963e-10
0.6956521739130435 4.018718691156664e-11
0.7507246376811594 6.344325065299472e-11
0.8057971014492753 6.568479093971291e-11
0.8608695652173912 1.6892709453486532e-11
0.9159420289855071 2.7922442136230075e-10
0.9710144927536233 7.806288948586371e-11
1.0260869565217392 2.495585960105018e-10
1.0811594202898551 1.4535439518681414e-10
1.136231884057971 1.9332859757525966e-09
1.191304347826087 1.0249023851827133e-10
1.2463768115942029 1.8081758312860075e-10
1.3014492753623188 6.23874285565762e-11
1.3565217391304347 2.346585148416125e-10
1.4115942028985506 3.5783820351298345e-11
1.4666666666666666 1.3962586642435326e-10
1.5217391304347825 5.809575043258519e-12
1.5768115942028984 2.732143400407949e-10
1.6318840579710143 2.3146107253069204e-10
1.6869565217391305 1.8059242989920676e-10
1.7420289855072464 6.216960279914474e-11
1.7971014492753623 4.075968451644485e-10
1.8521739130434782 1.9317880628477724e-10
1.9072463768115941 2.6530444507955053e-10
1.96231884057971 1.3339251925259532e-09
2.017391304347826 5.426323834711866e-10
2.072463768115942 1.2362999513015893e-10
2.127536231884058 1.3024759049073964e-10
2.1826086956521737 1.1393708199136654e-10
2.23768115942029 3.3777713959182165e-10
2.292753623188406 0.004456696985551067
2.347826086956522 8.042755350601283e-10
2.402898550724638 2.850291114242509e-10
2.457971014492754 1.4353673805089784e-10
2.5130434782608697 2.8158768650143884e-09
2.5681159420289856 7.1147017166595106e-09
2.6231884057971016 0.001411785106214758
2.6782608695652175 0.0011568820357095166
2.7333333333333334 2.6445134970742856e-10
2.7884057971014493 0.0007726018133940027
2.8434782608695652 9.434355519033488e-10
2.898550724637681 0.0005100158968727175
2.953623188405797 0.00041499390344212905
3.008695652173913 1.235412553146631e-06
3.063768115942029 0.0005427781442783886
3.118840579710145 0.0004351837625149457
3.173913043478261 0.00017565745724912318
3.228985507246377 4.9599566898095304e-08
3.284057971014493 0.00011283968600928063
3.339130434782609 2.4116286745368143e-10
3.394202898550725 7.172771155206092e-05
3.449275362318841 2.212257044220678e-10
3.5043478260869567 4.539377011769652e-05
3.5594202898550726 2.2837439761502765e-07
3.6144927536231886 5.705783188170166e-05
3.6695652173913045 1.9731223366292738e-07
3.7246376811594204 3.231552894167322e-07
3.7797101449275363 1.3987151343375359e-05
3.8347826086956522 7.325890583276262e-06
3.889855072463768 8.614697517783654e-06
3.944927536231884 6.8678610531947015e-06
4.0 2.6680439857429405e-08
                   };
        \end{axis}
\end{tikzpicture}
\qquad
\begin{tabular}[c]{lll}\hline
      Ansatze & \# CNOT & \# Parameter  \\ \hline
      $\mathbf{A_{4,2,s_z=0}}$ & \bf{6} & \bf{3} \\
      $\mathbf{E_{4,2}}$ & \bf{14} & \bf{5} \\
      $\mathbf{E_{4,2,s=0,s_z=0}}$ & \bf{20} & \bf{2} \\
      UCCSD & 56 & 3 \\
      SWAPRZ & 34 & 72 \\
      RYRZ & 18 & 32 \\
      RY & 18 & 16 \\\hline
    \end{tabular}
\raggedright
\caption{Energy difference from the exact ground state of $H_2$ for various ans\"atze. All ans\"atze perform well below chemical accuracy, but vary significantly in their number of required parameters and CNOT gates (shown in the accompanying table). Our proposed ansatze ($A_{4,2}$ and $E_{4,2}$) achieve very small energy differences and have low to modest CNOT counts. The UCCSD ansatz also performs very well with low parameters (3), but has a fairly large CNOT count in this example.}
\label{fig:diss1}
\end{figure*}

\begin{figure*}[!htb]
    \definecolor{mmaBlue}{HTML}{5e81b5}
    \definecolor{mmaOrange}{HTML}{e19c24}
    \definecolor{mmaGreen}{HTML}{8fb032}
    \definecolor{mmaRed}{HTML}{eb6235}
    \definecolor{mmaPurple}{HTML}{8778b3}
    \definecolor{mmaBrown}{HTML}{c56e1a}
    \definecolor{mmaBlack}{HTML}{696969}
    \raggedright
    \pgfplotsset{compat=newest}
    \begin{tikzpicture}[baseline=0.5 cm]
            \begin{axis}[
                xlabel = { Interatomic Distance [\AA]},
                ylabel = {$\Delta$E (Hartree)},
                xmin = 0,
                xmax = 4.0,
                ymin = 0.001,
                ymax = 0.35,
                ytick={0.05,0.1,0.15,0.2,0.25,0.3,0.35},
                y tick label style={/pgf/number format/.cd,scaled y ticks = false,
              set thousands separator={},
              fixed},
                legend style={at={(1,1)},anchor=north east, 
                legend columns =1}
                ]
            \addlegendentry{$A_{4,2,s_z=0}$}
            \addlegendentry{$E_{4,2}$}
            \addlegendentry{$E_{4,2,s=0,s_z=0}$}
            \addlegendentry{RY}
                 \addplot[line width=0.75pt,solid,color=mmaBlue,mark=*]
                    table{
                      0.3 0.046743749946818625
                      0.4 0.05578315709793902
                      0.5 0.04704495098075956
                      0.6000000000000001 0.05108084951402647
                      0.7000000000000002 0.03565201237057147
                      0.8000000000000003 0.04080747364953918
                      0.9000000000000001 0.026999253695722203
                      1.0000000000000002 0.028407296056627285
                      1.1000000000000003 0.03480339304293634
                      1.2000000000000004 0.028735271990422362
                      1.3000000000000005 0.02284495380775864
                      1.4000000000000004 0.016959284864885737
                      1.5000000000000004 0.02006099048690224
                      1.6000000000000005 0.022727763489983177
                      1.7000000000000004 0.021187458810038473
                      1.8000000000000005 0.014427949164526055
                      1.9000000000000006 0.019399194689731836
                      2.0000000000000004 0.01971203715480141
                      2.1000000000000005 0.0204642421724881
                      2.2000000000000006 0.021781440451933642
                      2.3000000000000007 0.022456936772304203
                      2.4000000000000004 0.024156642249793192
                      2.5000000000000004 0.021198157139614926
                      2.6000000000000005 0.022396021865513482
                      2.7000000000000006 0.022357638649108225
                      2.8000000000000007 0.02509778369909954
                      2.900000000000001 0.021308270556675835
                      3.000000000000001 0.020269782232550826
                      3.1000000000000005 0.02212871421684981
                      3.2000000000000006 0.022540556486892793
                      3.3000000000000007 0.02352733917579597
                      3.400000000000001 0.025264302343147582
                      3.500000000000001 0.023862291650973866
                      3.600000000000001 0.023124970869747186
                      3.700000000000001 0.025152060146488076
                      3.800000000000001 0.023223971453639525
                      3.9000000000000012 0.025957875377977446
                       };
                        \addplot[line width=0.75pt,solid,color=mmaGreen,mark=+]
                    table{
                        0.3	0.30512194
                        0.4	0.26771764
                        0.5	0.26839662
                        0.6	0.22489149
                        0.7	0.22925203
                        0.8	0.21246152
                        0.9	0.20145938
                        1	0.19888763
                        1.1	0.19715862
                        1.2	0.16917979
                        1.3	0.1607544
                        1.4	0.16029194
                        1.5	0.13305095
                        1.6	0.14502423
                        1.7	0.13220638
                        1.8	0.1049759
                        1.9	0.11529549
                        2	0.1104611
                        2.1	0.08650931
                        2.2	0.06900454
                        2.3	0.0623825
                        2.4	0.06446592
                        2.5	0.05887226
                        2.6	0.07277223
                        2.7	0.07398801
                        2.8	0.0525079
                        2.9	0.06412415
                        3	0.06102027
                        3.1	0.06552573
                        3.2	0.06026667
                        3.3	0.06404097
                        3.4	0.07473422
                        3.5	0.06636196
                        3.6	0.05838754
                        3.7	0.07023951
                        3.8	0.06618134
                        3.9	0.06706272
                       };
                       \addplot[line width=0.75pt,solid,color=mmaRed,mark=x]
                    table{
                    0.3	0.22722185
                    0.4	0.22723591
                    0.5	0.20060707
                    0.6	0.19081816
                    0.7	0.15851953
                    0.8	0.14957362
                    0.9	0.13664793
                    1	0.13104375
                    1.1	0.13482993
                    1.2	0.11891799
                    1.3	0.1142301
                    1.4	0.11421741
                    1.5	0.10407944
                    1.6	0.10542399
                    1.7	0.1073274
                    1.8	0.09700264
                    1.9	0.09681857
                    2	0.09916097
                    2.1	0.09751952
                    2.2	0.09498823
                    2.3	0.09574332
                    2.4	0.09465893
                    2.5	0.09421343
                    2.6	0.09156679
                    2.7	0.09223748
                    2.8	0.09101946
                    2.9	0.09235865
                    3	0.09458746
                    3.1	0.09428228
                    3.2	0.09416353
                    3.3	0.09164484
                    3.4	0.09081885
                    3.5	0.09413221
                    3.6	0.09318007
                    3.7	0.09408178
                    3.8	0.09287087
                    3.9	0.09294523
                    };
                    \addplot[line width=0.75pt,solid,color=mmaBlack,mark=pentagon]
                    table{
                    0.3	0.11953103
                    0.4	0.10094377
                    0.5	0.11631582
                    0.6	0.08891452
                    0.7	0.06349646
                    0.8	0.07504497
                    0.9	0.05314472
                    1	0.06275286
                    1.1	0.03976006
                    1.2	0.05273485
                    1.3	0.04046969
                    1.4	0.03172937
                    1.5	0.03237993
                    1.6	0.03599708
                    1.7	0.03327706
                    1.8	0.03762154
                    1.9	0.04498547
                    2	0.0349015
                    2.1	0.0390829
                    2.2	0.04610481
                    2.3	0.03257437
                    2.4	0.03781717
                    2.5	0.03626594
                    2.6	0.0320597
                    2.7	0.03620235
                    2.8	0.04019591
                    2.9	0.04135841
                    3	0.03669089
                    3.1	0.040579
                    3.2	0.04156157
                    3.3	0.04206506
                    3.4	0.03714949
                    3.5	0.0462838
                    3.6	0.04047381
                    3.7	0.04106233
                    3.8	0.04092854
                    3.9	0.0421896
                    };
                             \end{axis}
            \node[above right]
    at (current bounding box.south west) {a)};
        \end{tikzpicture}
    \begin{tikzpicture}[baseline=0.5 cm]
          \begin{axis}[
            axis y line*=left,
            xlabel = {Interatomic Distance [\AA]},
            ylabel = {$s$},
            xmin = 0,
            xmax = 4.0,
            ymin= 0,
            ymax= 1,
            xtick pos=left,
            ytick pos=left,
            ]
        \node at (15,5) {b)};
             \addplot[line width=0.75pt,solid,color=mmaBlue,mark=*]
                table{
                    0.3	0.04227307
                    0.4	0.039120785
                    0.5	0.04459986
                    0.6	0.036121659
                    0.7	0.040763588
                    0.8	0.043434618
                    0.9	0.046279095
                    1	0.051624836
                    1.1	0.057752927
                    1.2	0.0566873
                    1.3	0.038351318
                    1.4	0.041971071
                    1.5	0.086763831
                    1.6	0.046261965
                    1.7	0.049716739
                    1.8	0.05818839
                    1.9	0.103555223
                    2	0.051951483
                    2.1	0.471316259
                    2.2	0.063652014
                    2.3	0.428287371
                    2.4	0.372823149
                    2.5	0.035034609
                    2.6	0.255729952
                    2.7	0.924203363
                    2.8	0.94860669
                    2.9	0.925840484
                    3	0.928167511
                    3.1	0.950266819
                    3.2	0.958250368
                    3.3	0.949417567
                    3.4	0.875512116
                    3.5	0.781372326
                    3.6	0.769851658
                    3.7	0.959437726
                    3.8	0.91802922
                    3.9	0.451364237
};
                    \addplot[line width=0.75pt,solid,color=mmaGreen,mark=+]
                table{
                    0.3	0.130291727
                    0.4	0.169368548
                    0.5	0.165513426
                    0.6	0.181871439
                    0.7	0.169115753
                    0.8	0.349038808
                    0.9	0.210851004
                    1	0.171185254
                    1.1	0.158958303
                    1.2	0.208213494
                    1.3	0.329628307
                    1.4	0.170020688
                    1.5	0.488623645
                    1.6	0.867984793
                    1.7	0.878264152
                    1.8	0.851490696
                    1.9	0.885889113
                    2	0.866135953
                    2.1	0.869420303
                    2.2	0.882391001
                    2.3	0.891108391
                    2.4	0.892588878
                    2.5	0.884801844
                    2.6	0.879908158
                    2.7	0.877296959
                    2.8	0.882325461
                    2.9	0.877275541
                    3	0.914814468
                    3.1	0.869283769
                    3.2	0.877622231
                    3.3	0.893204981
                    3.4	0.896262841
                    3.5	0.919037751
                    3.6	0.890473012
                    3.7	0.737464751
                    3.8	0.886938832
                    3.9	0.875454726
                   };
                   \addplot[line width=0.75pt,solid,color=mmaRed,mark=x]
                table{
                0.3	0.083774436
                0.4	0.084794044
                0.5	0.082976216
                0.6	0.083152707
                0.7	0.088959352
                0.8	0.086538713
                0.9	0.095336016
                1	0.086664937
                1.1	0.093674415
                1.2	0.091382319
                1.3	0.088860453
                1.4	0.094838158
                1.5	0.092563309
                1.6	0.08980746
                1.7	0.093548504
                1.8	0.090601405
                1.9	0.0986198
                2	0.097614465
                2.1	0.095853851
                2.2	0.098844339
                2.3	0.098409159
                2.4	0.102361982
                2.5	0.100602171
                2.6	0.103371581
                2.7	0.084507449
                2.8	0.10407327
                2.9	0.103758043
                3	0.104324382
                3.1	0.085078731
                3.2	0.104476612
                3.3	0.103940805
                3.4	0.102986959
                3.5	0.104398728
                3.6	0.084988261
                3.7	0.102656224
                3.8	0.104038748
                3.9	0.105024669
                };
                \addplot[line width=0.75pt,solid,color=mmaBlack,mark=pentagon]
                table{
                0.3	0.048331031
                0.4	0.052242
                0.5	0.060900939
                0.6	0.049646497
                0.7	0.045155968
                0.8	0.052337696
                0.9	0.068365571
                1	0.105642359
                1.1	0.071008349
                1.2	0.036575092
                1.3	0.147745523
                1.4	0.438420115
                1.5	0.096295775
                1.6	0.959728171
                1.7	0.955197679
                1.8	0.960790216
                1.9	0.958196132
                2	0.040309371
                2.1	0.964262393
                2.2	0.969223751
                2.3	0.958024314
                2.4	0.958391894
                2.5	0.977220553
                2.6	0.969866155
                2.7	0.954451716
                2.8	0.968773097
                2.9	0.968415924
                3	0.967316304
                3.1	0.95752001
                3.2	0.961748627
                3.3	0.952969836
                3.4	0.967201032
                3.5	0.967887001
                3.6	0.968775743
                3.7	0.953282683
                3.8	0.953903799
                3.9	0.951917312
                };
        \end{axis}
        \begin{axis}[
            ymode=log,
            axis y line*=right,
            axis x line= none,
            xtick pos=right,
            ytick pos=right,
            ylabel = {Triplet-Singlet Energy Difference}
            ]
            \addplot[line width=0.75pt,dashed,color=mmaOrange,mark=none]
            table{
            0.3 1.437879310830926
            0.4 1.1970838747814416
            0.5 0.9844197643341035
            0.6000000000000001 0.8053259502992632
            0.7000000000000002 0.6577364476990046
            0.8000000000000003 0.5369699116400437
            0.9000000000000001 0.4377109267793444
            1.0000000000000002 0.3552785858471349
            1.1000000000000003 0.28623329548295695
            1.2000000000000004 0.22829744687091846
            1.3000000000000005 0.17994936786616422
            1.4000000000000004 0.14004034601461035
            1.5000000000000004 0.10756459989951561
            1.6000000000000005 0.08156093595576164
            1.7000000000000004 0.06108927610306236
            1.8000000000000005 0.04524206384698348
            1.9000000000000006 0.03316913385162201
            2.0000000000000004 0.024103803203825214
            2.1000000000000005 0.017381997605069532
            2.2000000000000006 0.012450404766227718
            2.3000000000000007 0.008863752740134734
            2.4000000000000004 0.006274084826690229
            2.5000000000000004 0.004415836124373795
            2.6000000000000005 0.0030899329167652922
            2.7000000000000006 0.0021490730356839016
            2.8000000000000007 0.0014852375759484282
            2.900000000000001 0.0010197012450119392
            3.000000000000001 0.0006953518282228899
            3.1000000000000005 0.0004709288873556705
            3.2000000000000006 0.0003167612144308052
            3.3000000000000007 0.0002116298956067464
            3.400000000000001 0.00014046136379253227
            3.500000000000001 9.262928047126806e-05
            3.600000000000001 6.0704927974386536e-05
            3.700000000000001 3.954028156338829e-05
            3.800000000000001 2.5599080780480676e-05
            3.9000000000000012 1.6472972483416726e-05
}; 
\label{diff_plot}
        \end{axis}
        \node[above right]
    at (current bounding box.south west) {b)};
    \end{tikzpicture}
\caption{a) Energy difference from the exact ground state of $H_2$ in the presence of noise characterized by IBM's Poughkeepsie device. Included is Qiskit's state preperation and measurement error mitigation for all ans\"atze. b) Total spin  eigenvalues for the same dissociation curve (left axis). The total spin values indicate that some ans\"atze drift outside the singlet subspace ($s=0$) and begin to instead find a (nearly degenerate) triplet state ($s=1$) at large bond distances. Since the $E_{4,2,s=0,s_z=0}$ ansatz is restricted to a fixed total spin, it is the only displayed ans\"atze which always attempts to find the true ground state across the full dissociation curve. The energy difference between this triplet state and the true singlet ground state as a function of interatomic distance is also shown for comparison (orange dashed line, right axis). }
\label{fig:diss_noise}
\end{figure*}
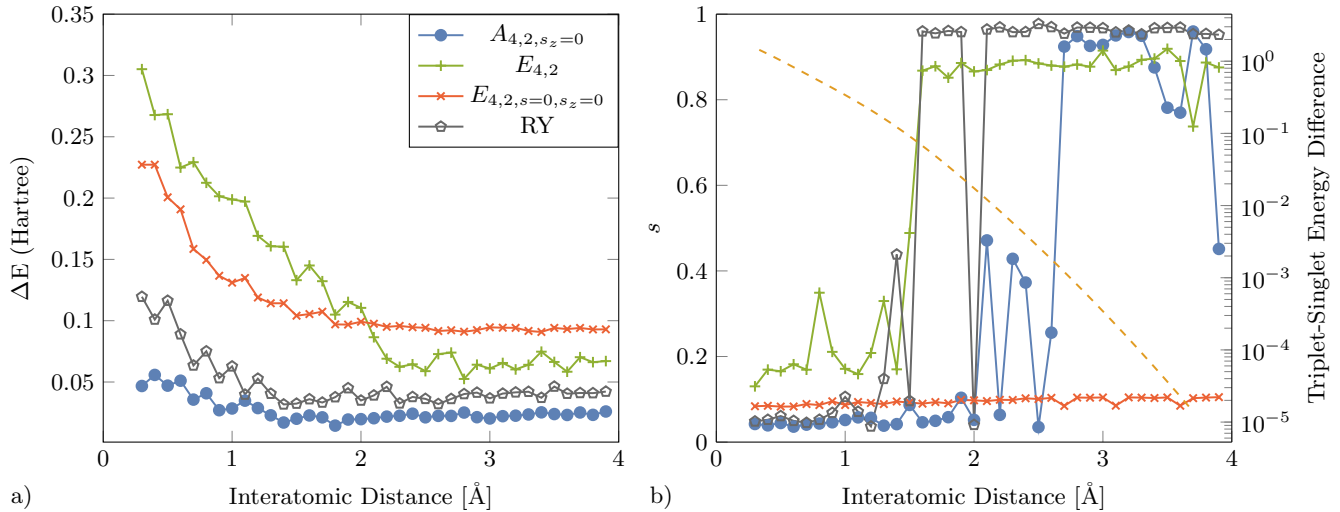
\begin{figure*}[!ht]
    \definecolor{mmaBlue}{HTML}{5e81b5}
    \definecolor{mmaOrange}{HTML}{e19c24}
    \definecolor{mmaGreen}{HTML}{8fb032}
    \definecolor{mmaRed}{HTML}{eb6235}
    \definecolor{mmaPurple}{HTML}{8778b3}
    \definecolor{mmaBrown}{HTML}{c56e1a}
    \definecolor{mmaBlack}{HTML}{696969}
    \centering
    \pgfplotsset{compat=newest}
    \begin{tikzpicture}[baseline=2.8 cm]
            \begin{axis}[
                xlabel = { Interatomic Distance [\AA]},
                ylabel = {$\Delta$E (Hartree)},
                xmin = 0.4,
                xmax = 3.6,
                ymin = 0.001,
                ymax = 0.32,
                ytick={0.05,0.1,0.15,0.2,0.25,0.3,0.35},
                y tick label style={/pgf/number format/.cd,scaled y ticks = false,
              set thousands separator={},
              fixed},
                legend style={at={(1,1)},anchor=north east, 
                legend columns =1}
                ]
            \addlegendentry{$A_{6,2,s_z=0}$}
            \addlegendentry{$E_{6,2,s=0,s_z=0}$}
            \addlegendentry{RY}
                 \addplot[line width=0.75pt,solid,color=mmaBlue,mark=*]
                    table{
                        0.5	 0.026087669
                        0.55	0.020959574
                        0.6	0.023078095
                        0.65	0.027422814
                        0.7	0.022210044
                        0.75	0.028723047
                        0.8	0.024089382
                        0.85	0.024281279
                        0.9	0.025195742
                        0.95	0.023536379
                        1	0.028210765
                        1.05	0.023595587
                        1.1	0.029046819
                        1.15	0.028195624
                        1.2	0.031076524
                        1.25	0.027092838
                        1.3	0.025332086
                        1.35	0.03145233
                        1.4	0.02653177
                        1.45	0.028574847
                        1.5	0.027045957
                        1.55	0.032203372
                        1.6	0.033776875
                        1.65	0.030032893
                        1.7	0.028913975
                        1.75	0.030644134
                        1.8	0.029036531
                        1.85	0.036722814
                        1.9	0.037014089
                        1.95	0.034560897
                        2	0.038278132
                        2.05	0.03931529
                        2.1	0.036850504
                        2.15	0.0435003
                        2.2	0.035494182
                        2.25	0.045099164
                        2.3	0.048252424
                        2.35	0.053063127
                        2.4	0.049693552
                        2.45	0.053327057
                        2.5	0.058141544
                        2.55	0.057649528
                        2.6	0.05785418
                        2.65	0.063342371
                        2.7	0.066660771
                        2.75	0.07164103
                        2.8	0.068731407
                        2.85	0.080364599
                        2.9	0.049810941
                        2.95	0.042399779
                        3	0.034422937
                        3.05	0.052208008
                        3.1	0.064270975
                        3.15	0.047174105
                        3.2	0.041075397
                        3.25	0.059129947
                        3.3	0.053322339
                        3.35	0.065926253
                        3.4	0.066321853
                        3.45	0.019306339
                        3.5	0.04576435
                        3.55	0.059656879
                       };
                        \addplot[line width=0.75pt,solid,color=mmaRed,mark=x]
                    table{
                        0.5	0.180905545
                        0.55	0.178543834
                        0.6	0.18354008
                        0.65	0.202827125
                        0.7	0.201247102
                        0.75	0.209407606
                        0.8	0.216929435
                        0.85	0.217400301
                        0.9	0.219353665
                        0.95	0.224559147
                        1	0.22843433
                        1.05	0.222883035
                        1.1	0.212431991
                        1.15	0.221168443
                        1.2	0.224258003
                        1.25	0.223124471
                        1.3	0.218973575
                        1.35	0.219720569
                        1.4	0.213414222
                        1.45	0.212754967
                        1.5	0.211416063
                        1.55	0.213713954
                        1.6	0.209282266
                        1.65	0.20447985
                        1.7	0.203778752
                        1.75	0.202912882
                        1.8	0.200581306
                        1.85	0.197647994
                        1.9	0.190979926
                        1.95	0.195793823
                        2	0.197191621
                        2.05	0.192401185
                        2.1	0.193427635
                        2.15	0.189491069
                        2.2	0.193083266
                        2.25	0.186698576
                        2.3	0.189489509
                        2.35	0.18987017
                        2.4	0.191252793
                        2.45	0.194731958
                        2.5	0.190955992
                        2.55	0.190124436
                        2.6	0.196288261
                        2.65	0.183854301
                        2.7	0.193401316
                        2.75	0.187936122
                        2.8	0.195632261
                        2.85	0.196821653
                        2.9	0.191465357
                        2.95	0.194316953
                        3	0.204545279
                        3.05	0.205522772
                        3.1	0.207099086
                        3.15	0.206918338
                        3.2	0.209252228
                        3.25	0.2132656
                        3.3	0.207667241
                        3.35	0.219433928
                        3.4	0.217787699
                        3.45	0.221733358
                        3.5	0.223816608
                        3.55	0.226348583
                       };
                    \addplot[line width=0.75pt,solid,color=mmaBlack,mark=pentagon]
                    table{
                    0.5	0.114128226
                    0.55	0.1255775
                    0.6	0.123855484
                    0.65	0.123764581
                    0.7	0.124681432
                    0.75	0.119200955
                    0.8	0.117364122
                    0.85	0.121397116
                    0.9	0.126830763
                    0.95	0.121448498
                    1	0.133760123
                    1.05	0.113814098
                    1.1	0.127613499
                    1.15	0.123569073
                    1.2	0.12132449
                    1.25	0.138475205
                    1.3	0.133862196
                    1.35	0.125651251
                    1.4	0.117789691
                    1.45	0.127849714
                    1.5	0.12534156
                    1.55	0.124127129
                    1.6	0.124003605
                    1.65	0.117952508
                    1.7	0.119859434
                    1.75	0.116975023
                    1.8	0.118743257
                    1.85	0.119119637
                    1.9	0.119041997
                    1.95	0.120167921
                    2	0.112215444
                    2.05	0.12176935
                    2.1	0.115487554
                    2.15	0.121178287
                    2.2	0.121640966
                    2.25	0.124757567
                    2.3	0.116132704
                    2.35	0.113119408
                    2.4	0.122096493
                    2.45	0.118060238
                    2.5	0.114775468
                    2.55	0.11843329
                    2.6	0.128009701
                    2.65	0.132985647
                    2.7	0.133656793
                    2.75	0.139882666
                    2.8	0.140522793
                    2.85	0.141681903
                    2.9	0.146323982
                    2.95	0.144343346
                    3	0.132843246
                    3.05	0.131232785
                    3.1	0.133939582
                    3.15	0.132502179
                    3.2	0.123777099
                    3.25	0.118051011
                    3.3	0.112287016
                    3.35	0.102415563
                    3.4	0.116504743
                    3.45	0.093514323
                    3.5	0.099131345
                    3.55	0.105448701
                };
                             \end{axis}
        \end{tikzpicture}
        \qquad
\begin{tabular}[c]{lll}\hline
      Ansatze & \# CNOT & \# Parameter  \\ \hline
      $\mathbf{A_{6,2,s_z=0}}$ & \bf{25} & \bf{8} \\
      $\mathbf{E_{6,2,s=0,s_z=0}}$ & \bf{48} & \bf{5} \\
      RY & 5 & 12 \\\hline
    \end{tabular}
\raggedright
        \caption{Energy difference from the exact ground state of $LiH$ in the presence of noise characterized by IBM's Poughkeepsie device. With appropriate orbital reductions, $LiH$ is run on 6 qubits and has two excitations. The A-gate ansatz continues to perform well utilizing particle number and spin projection symmetries and relatively low CNOT count for this dimension.}
        \label{fig:LiH}
\end{figure*}

To demonstrate the efficiency of our state preparation circuits, we use them in a VQE simulation to compute the ground state energy of the $H_2$ molecule. We work in the STO-3G basis and map to qubits using the Jordan-Wigner transform. 
We note that there are many choices of qubit mappings and unitary transformations that can be used to reduce the complexity of molecular simulations~\cite{Bravyi2017}. However, we have formulated our ans\"atze to preserve symmetries in the natural basis where each qubit represents a physical spin orbital. There are also several techniques to tailor an ansatz to efficiently describe the allowed transitions in $H_2$, but here we are concerned with constructing general ans\"atze, applicable to any molecule, which only utilize particle number and spin symmetries, to reduce the complexity of the resulting circuits. Incorporation of particle number, spin projection and total spin symmetries in conjuction with other qubit mappings and reduction methods is an interesting topic, but beyond the scope of the present work.
The results are given in Fig.~\ref{fig:diss1}, which shows the difference between the computed ground state energy and the exact ground state energy. For comparison, we also show the results obtained from standard ans\"atze (SWAPRZ, RY, RYRZ, and UCCSD) included in IBM's Qiskit software package~\cite{Qiskita}. Furthermore, we show the variational parameter and CNOT gate counts in the table accompanying Fig.~\ref{fig:diss1} for all ans\"atze considered. It is clear that our proposed circuits obtain the ground state energy with high accuracy using the minimal number of variational parameters and a low number of CNOT gates. For the RY, RYRZ, and SWAPRZ ans\"atze, we choose a circuit depth of three, while for UCCSD, we use a circuit depth of one. In all cases, we use the BFGS optimizer (run several times with random initial conditions) included in Qiskit and consider an ideal, noiseless simulation.

Fig.~\ref{fig:diss_noise} shows the results of a similar simulation, but this time with noise included. This noise is characterized by the physical parameters of IBM's Poughkeepsie device. Here we only take the standard ansatz that has reasonable parameter and CNOT scaling, RY, and compare it to our ans\"atze. Contrary to the noiseless case, the RY ansatz here is chosen to have linear connectivity (nearest-neighbor CNOT connectivity) and a circuit depth of one as this configuration performs best under these noisy conditions. With these settings, the RY ansatz has 8 parameters and 3 CNOT gates. The CNOT and parameter counts for our ansatze are the same as in Fig.~\ref{fig:diss1} since these are determined by the ground state symmetries of the molecule. Even though the RY ansatz has fewer costly CNOT gates than our $A$ gate ansatz and is over parameterized for the target ground state, it is outperformed by the $A$ gate ansatz. This is due to the fact that at this low depth, the RY ansatz does not have sufficient coverage of the target Hilbert space. Increasing the depth of the RY ansatz will increase its coverage of the target space, but also increases the number of CNOT gates. This is not an equivalent trade-off and actually decreases the ability of the ansatz to approximate the ground state. In the case of both $E$ gate ans\"atze, their relatively larger number of CNOT gates can be seen to hinder their performance. However, as these gates are constructed in terms of arbitrary state generation, these gate counts could be improved with more efficient state generation algorithms. In addition, mitigation of CNOT errors may be possible with the use of Richardson extrapolation~\cite{Li2017,McCaskey2019}, a technique we will pursue in future work.

An interesting feature in Fig.~\ref{fig:diss_noise} a) can be seen where the $E_{4,2}$ ansatz crosses the $E_{4,2,s=0,s_z=0}$ ansatz. In principle, the spin symmetry ansatz should always have a better ability to target the ground state in a VQE algorithm.  This is explained in Fig.~\ref{fig:diss_noise} b), where as we see that the $E_{4,2}$ ansatz abruptly changes the total spin value of its lowest energy state. Indeed, the triplet states ($s=1$) are nearly degenerate with the singlet state ($s=0$, Eq.~\ref{eq:s0sz0}) at large bound distances. At an interatomic distance near two Angstroms, the energy difference between the lowest energy singlet and triplet states is $\mathcal{O}(10^{-2})$ Hartree. The optimization landscape becomes much more difficult to navigate when there are many nearly equal local minima, as is the case here at large bond distances. Therefore since the algorithm attempts to find the lowest energy eigenvalue and the singlet and triplet states are nearly degenerate at longer bond distances, the optimizer will frequently return a minimum (triplet state) rather than the global minimum (singlet state). In the presence of noise, this issue is blurred further as both energies are not exactly attainable, but a triplet state is trivial to produce (e.g. $s=1,s_z=1$, $\ket{1010}$), so finding the triplet state as the lowest energy state is frequently encountered. Additionally, this issue grows more difficult to navigate with larger molecules and strongly correlated systems, which have many more nearly degenerate ground states. Naturally one could also enforce a spin eigenvalue constraint on the optimizer by modifying the objective function, but this places stricter requirements on the optimizer and requires a physical measurement of spin eigenvalues, through tomography or additional circuit elements. However, enforcing symmetry in the ans\"atze lessens the burden on the optimizer and does not directly require spin measurements.

We also perform similar noisy simulations on $LiH$. $LiH$ in the STO-3G basis maps to 12 qubits but we follow the same methods as Ref.~\cite{Kandala2017}, removing two non-interacting orbitals and freezing the core orbital, reducing $LiH$ to 6 qubits. We show in Fig.~\ref{fig:LiH} that our ans\"atze perform comparably for $LiH$ as they do for $H_2$ with notably more error for the $E$ gate ansatz as it requires significantly more CNOTs. As in the noisy $H_2$ case, the RY ansatz is chosen to have a fixed depth (one) and connectivity (nearest-neighbor). These settings give the best performance under these noisy conditions but the same coverage-CNOT trade off issue arises as before. From both the $H_2$ and $LiH$ results we can see that the $A$-gate circuit finds a middle-ground in terms of CNOT and parameter count and has good performance, while the $E$-gate circuit exactly minimizes parameter count based on particle number, spin projection, and total spin, but its performance suffers due to a large CNOT count. Thus, in the presence of noise, we find that it can be favorable to relax some of the symmetry constraints to reduce the circuit depth at the expense of more variational parameters.

Since our constructions maintain number and spin symmetries, potentially any noise sources which violate these symmetries can be mitigated through post-selection or symmetry verification~\cite{Bonet2018,McArdle2018,Sagastizabal2019}. This will be a topic of future work. Another interesting question for future work is whether it is possible to find circuits that minimize the number of CNOT gates while respecting spin symmetries and while using the minimal number of parameters necessary to span the symmetry subspace.

\section{Discussion}\label{sec:conclusion}
In this work, we presented general schemes to construct state preparation circuits for quantum simulation that respect a number of symmetries that commonly arise in systems of interest, including particle number, time-reversal, total spin, and spin magnetization. In each case, we provide general construction procedures, explicit examples of circuits, and gate counts. In the case of particle-number symmetry, for which state preparation circuits have been proposed previously by other authors, our circuits outperform existing methods in terms of the number of two-qubit entangling gates they contain. Enforcing spin symmetries in addition to particle number can significantly enhance the performance of the variational quantum eigensolver by preventing the algorithm from wasting time exploring vast, irrelevant regions of Hilbert space.
Also, our ans\"atze are \textit{guaranteed} to contain the ground state of interest, with the minimal number of parameters defined by the symmetry subspace.
It is important to emphasize, however, that the number of variational parameters still grow exponentially with system size even when all symmetries are imposed. Thus, spanning the entire symmetry subspace will become impractical as the size of the quantum processor is increased beyond a few tens of qubits. However, we expect that our symmetry preservation techniques will continue to play an important role for larger NISQ devices as systematic methods to further reduce the search to still smaller regions of Hilbert space are developed. This could be done, for example, by looking for ways to freeze large sets of the variational parameters appearing in our circuits without sacrificing the accuracy of the ansatz.
In addition, as a general state creation circuit with a particular symmetry, one can imagine applications for these circuits in the Phase Estimation Algorithm (PEA).

\section{Methods}
Confirmation of circuit compilation was performed in Mathematica software as well as numerical optimization to confirm that the proposed circuits span any state in the desired Hilbert space. In numerical optimization testing, we find that Mathematica's NMaximize method ``Simulated Annealing'' performs best for our purposes, when run over many samples of a randomly chosen state in the Hilbert space. Gate decomposition was performed using the method of Gray codes, was done ``by hand'' and confirmed in Mathematica.

Our VQE simulations were done using IBM's Qiskit software, coding in our own proposed ans\"atze, which enforce relevant symmetries. Noisy simulations were simulated using the noise parameters of IBM's Poughkeepsie device.

\section{Data Availability}
The data that support the findings of this study are available from the authors upon request.

\section{Code Availability}
A custom Mathematica code to reproduce our results is available on GitHub~\cite{Gardgit2019a}.

\section*{Acknowledgements}
This research was supported by the US Department of Energy (Award No. de-sc0019199) and the National Science Foundation (Award No. 1839136).  S.E.E. also acknowledges support from Award No. de-sc0019318 from the Department of Energy. This research used quantum computing system resources supported by the U.S. Department of Energy, Office of Science, Office of Advanced Scientific Computing Research program office. Oak Ridge National Laboratory manages access to the IBM Q System as part of the IBM Q Network.

\section{Author Contributions}
S.E.E., E.B., and N.J.M. structured and supervised the project. E.B. proposed methods for symmetry encoding and minimal parameterization. B.T.G. performed simulations, formulated general circuits and drafted the manuscript. L.Z. performed circuit decomposition, CNOT gate counts and suggested example circuits. G.S.B. drafted figures and confirmed circuit compilations. All authors contributed to the revisions of the manuscript.

\section{Competing interests}
The authors declare that there are no competing interests.

\begin{table*}[!p]
    \begin{tabular}{|c|c|c|c|c|c|}
    \hline
         ${s,s_z}$ & $u_1$ & $u_2$ & $u_3$ & $u_4$ & $u_5$  \\
         \hline
         {1,0} & $\pi/2$ & $\pi/2$ & $\pi/4$ & $\pi/2$ & 0\\
         \hline
         {0,0} & $u_1$ & $\pi/2$ & $-\tan^{-1}(\csc u_4)$ & $u_4$ & $0$\\
         \hline
    \end{tabular}
    \caption{By fixing the coefficients of Eq.~\ref{eq:egateaction}, we can generate the two spin subspaces defined by their quantum numbers for the $n=4,m=2$ space. In the case of $s=1$, $s_z=0$, the subspace is spanned by a single state, so all the $u_i$ are fixed. The subspace with $s=s_z=0$ is three-dimensional, and so two of the $u_i$ are left unspecified.}
    \label{tbl1}
\end{table*}

\begin{table*}[!htb]
    \begin{tabular}{|c|c|c|c|c|c|c|c|c|c|c|c|c|c|c|c|c|c|c|c|}
    \hline
         ${s,s_z}$ & $u_1$ & $u_2$ & $u_3$ & $u_4$ & $u_5$ & $u_6$ & $u_7$ & $u_8$ & $u_9$ & $u_{10}$ & $u_{11}$ & $u_{12}$ & $u_{13}$ & $u_{14}$ & $u_{15}$ & $u_{16}$ & $u_{17}$ & $u_{18}$ & $u_{19}$  \\
         \hline
         $\frac{3}{2},\frac{1}{2}$ & $\frac{\pi }{2}$ & $\frac{\pi }{2}$ & $\frac{\pi }{2}$ & $\frac{\pi }{2}$ & $\cot^{-1}(\frac{1}{\sqrt{2}})$ & $\frac{\pi }{2}$ & $\frac{\pi }{2}$ & $\frac{\pi }{4}$ & $\frac{\pi }{2}$ & $\frac{\pi }{2}$ & $\frac{\pi }{2}$ & $\frac{\pi }{2}$ & $\frac{\pi }{2}$ & $\frac{\pi }{2}$ & 0 & $\frac{\pi }{2}$ & $\frac{\pi }{2}$ & $\frac{\pi }{2}$ & $\frac{\pi }{2}$\\
         \hline
         $\frac{3}{2},-\frac{1}{2}$ &
 $\frac{\pi }{2}$ & $\frac{\pi }{2}$ & $\frac{\pi }{2}$ & $\frac{\pi }{2}$ & $\frac{\pi }{2}$ & $\frac{\pi }{2}$ & $ \frac{\pi }{2}$ & $\frac{\pi }{2}$ & $ \cot^{-1}(\frac{1}{\sqrt{2}})$ & $\frac{\pi }{2}$ & $\frac{\pi }{2}$ & $\frac{\pi }{2} $& $\frac{\pi }{4}$ & $\frac{\pi }{2}$ & $\frac{\pi }{2}$ & 0 & $\frac{\pi }{2}$ & $\frac{\pi }{2}$ & $\frac{\pi }{2}$ \\
         \hline
 $\frac{1}{2},\frac{1}{2}$& $u_1$ & $\frac{\pi }{2}$ & $u_3$ & $\frac{\pi }{2}$ & $-\cot ^{-1}(\delta)$ &  $u_6$ & $\frac{\pi }{2}$ & $u_8$ & $\frac{\pi }{2}$ & $u_{10}$ & $\frac{\pi }{2}$ & $\frac{\pi }{2}$& $\frac{\pi }{2}$ & $\frac{\pi }{2}$ & $u_{15}$ & $\frac{\pi }{2}$ & $\frac{\pi }{2}$ & $u_{18}$ & 0 \\
 \hline
 $\frac{1}{2},-\frac{1}{2}$ & $\frac{\pi }{2}$ & $u_2$ & $\frac{\pi }{2}$ & $u_4$ & $\frac{\pi }{2}$ & $\frac{\pi }{2}$ & $\frac{\pi }{2}$ & $\frac{\pi }{2}$ & $-\cot^{-1}(\kappa)$ & $\frac{\pi }{2}$ & $u_{11}$ & $\frac{\pi }{2}$ & $u_{13}$ & $u_{14}$ & $\frac{\pi }{2}$ & $u_{16}$ & $u_{17}$ & $\frac{\pi }{2}$ & $\frac{\pi }{2}$ \\
\hline
    \end{tabular}
    \caption{By fixing the coefficients of Eq.~\ref{eq:fgate}, we can generate the four spin subspaces defined by their spin quantum numbers for the $n=6,~m=3$ space. Here $\delta=\{\sin (u_{10}) \cos (u_{15}) \sin (u_6) \sin (u_8)+\sin (u_6) \cos (u_8)\},~\kappa=\{\sin (u_{11}) \sin (u_{13}) \sin (u_{14}) \cos (u_{16})+\sin (u_{11}) \cos (u_{13})\}$. The order of basis states for this space is defined in the text.}
    \label{tbl2}
\end{table*}

\raggedright
\input{main1SI2.tex}
\end{document}

%% file: main1SI2.tex
\beginsupplement
\section{Supplemental Information}
\subsection{Construction of $E$ gate circuits}
As discussed in the main text, in order to find circuits that respect total spin, we start by constructing unitaries that transform $\otimes_{i=1}^n\ket{0}$ into an arbitrary state in a particular particle-number subspace, where the coefficients in this state are specified by hyperspherical coordinates (see Eq.~(5) of the main text). A specific example of such a unitary in the case of two fermions occupying four spin-orbitals is given by
\setcounter{MaxMatrixCols}{16}
\begin{widetext}
\begin{equation}\tag{S1}
E_4 =
\left(
\begin{array}{cccccccccccccccc}
 0 & 0 & 0 & 0 & 0 & 0 & 0 & 0 & 0 & 0 & 0 & 0 & 1 & 0 & 0 & 0 \\
 0 & 0 & 0 & 0 & 0 & 0 & 0 & 0 & 0 & 0 & 0 & 0 & 0 & 1 & 0 & 0 \\
 0 & 0 & 0 & 0 & 0 & 0 & 0 & 0 & 0 & 0 & 0 & 0 & 0 & 0 & 1 & 0 \\
 \text{C}_4 \text{S}_1 \text{S}_2 \text{S}_3 & 0 & 0 & 0 & 0 & \text{S}_4 & -\text{C}_1 \text{C}_4 \text{S}_2 \text{S}_3 & 0 & 0 & 0 & -\text{C}_2 \text{C}_4 \text{S}_3 & 0 & 0 & 0 & 0 & -\text{C}_3 \text{C}_4 \\
 0 & 0 & 0 & 0 & 0 & 0 & 0 & 0 & 1 & 0 & 0 & 0 & 0 & 0 & 0 & 0 \\
 \text{S}_1 \text{S}_2 \text{S}_3 \text{S}_4 \text{S}_5 & 0 & 0 & 0 & 0 & -\text{C}_4 \text{S}_5 & -\text{C}_1 \text{S}_2 \text{S}_3 \text{S}_4 \text{S}_5 & 0 & 0 & -\text{C}_5 & -\text{C}_2 \text{S}_3 \text{S}_4 \text{S}_5 & 0 & 0 & 0 & 0 & -\text{C}_3 \text{S}_4 \text{S}_5 \\
 \text{C}_3 \text{S}_1 \text{S}_2 & 0 & 0 & 0 & 0 & 0 & -\text{C}_1 \text{C}_3 \text{S}_2 & 0 & 0 & 0 & -\text{C}_2 \text{C}_3 & 0 & 0 & 0 & 0 & \text{S}_3 \\
 0 & 0 & 0 & 0 & 0 & 0 & 0 & 0 & 0 & 0 & 0 & 1 & 0 & 0 & 0 & 0 \\
 0 & 0 & 0 & 0 & 1 & 0 & 0 & 0 & 0 & 0 & 0 & 0 & 0 & 0 & 0 & 0 \\
 \text{C}_5 \text{S}_1 \text{S}_2 \text{S}_3 \text{S}_4 & 0 & 0 & 0 & 0 & -\text{C}_4 \text{C}_5 & -\text{C}_1 \text{C}_5 \text{S}_2 \text{S}_3 \text{S}_4 & 0 & 0 & \text{S}_5 & -\text{C}_2 \text{C}_5 \text{S}_3 \text{S}_4 & 0 & 0 & 0 & 0 & -\text{C}_3 \text{C}_5 \text{S}_4 \\
 \text{C}_2 \text{S}_1 & 0 & 0 & 0 & 0 & 0 & -\text{C}_1 \text{C}_2 & 0 & 0 & 0 & \text{S}_2 & 0 & 0 & 0 & 0 & 0 \\
 0 & 0 & 0 & 0 & 0 & 0 & 0 & 1 & 0 & 0 & 0 & 0 & 0 & 0 & 0 & 0 \\
 \text{C}_1 & 0 & 0 & 0 & 0 & 0 & \text{S}_1 & 0 & 0 & 0 & 0 & 0 & 0 & 0 & 0 & 0 \\
 0 & 1 & 0 & 0 & 0 & 0 & 0 & 0 & 0 & 0 & 0 & 0 & 0 & 0 & 0 & 0 \\
 0 & 0 & 1 & 0 & 0 & 0 & 0 & 0 & 0 & 0 & 0 & 0 & 0 & 0 & 0 & 0 \\
 0 & 0 & 0 & 1 & 0 & 0 & 0 & 0 & 0 & 0 & 0 & 0 & 0 & 0 & 0 & 0 \\
\end{array}
\right)
,\label{eq:E4gate}
\end{equation}
\end{widetext}
where $\hbox{C}_i=\cos{u_i}, \hbox{S}_i=\sin{u_i}$. The first column of this matrix is designed to satisfy Eq.~(5) of the main text, while the remaining columns are chosen to ensure unitarity. In Fig.~\ref{fig:e-gate}, we show an explicit symbolic decomposition of this unitary in terms of $4$-qubit Toffoli gates, which have a known decomposition into elementary gates~\cite{Barenco1995}. In general, a single $4$-qubit Toffoli gate can be decomposed into $13$ CNOT's~\cite{szyprowski2013}. However, it is important to note that for all pairs of identical $4$-qubit Toffoli gates, which occur frequently in our decomposition, we can utilize the approximate decompositions described by Barenco~\textit{et. al.}~\cite{Barenco1995}. Each of these pairs can be decomposed \textit{exactly} (the relative minus signs cancel due to the fact that we are decomposing pairs of identical 4-qubit Toffoli gates) into $18$ CNOT's, rather than the $26$ that result from the standard decomposition. This means that, in total, the $E_4$ gate can be decomposed into, at most, $155$ CNOT gates. In the most basic example of simplification (for the $\ket{s=1,s_z=0}$ case), the CNOT gate count can be as little as $67$ CNOT's. Our decomposition utilizes Gray codes, but as the 4-qubit gate itself is sparse, this is fairly efficient if one desires a symbolic decomposition. We also note that each $n$-bit Toffoli decomposition to elementary gates only grows polynomially with $n$~\cite{szyprowski2013}.

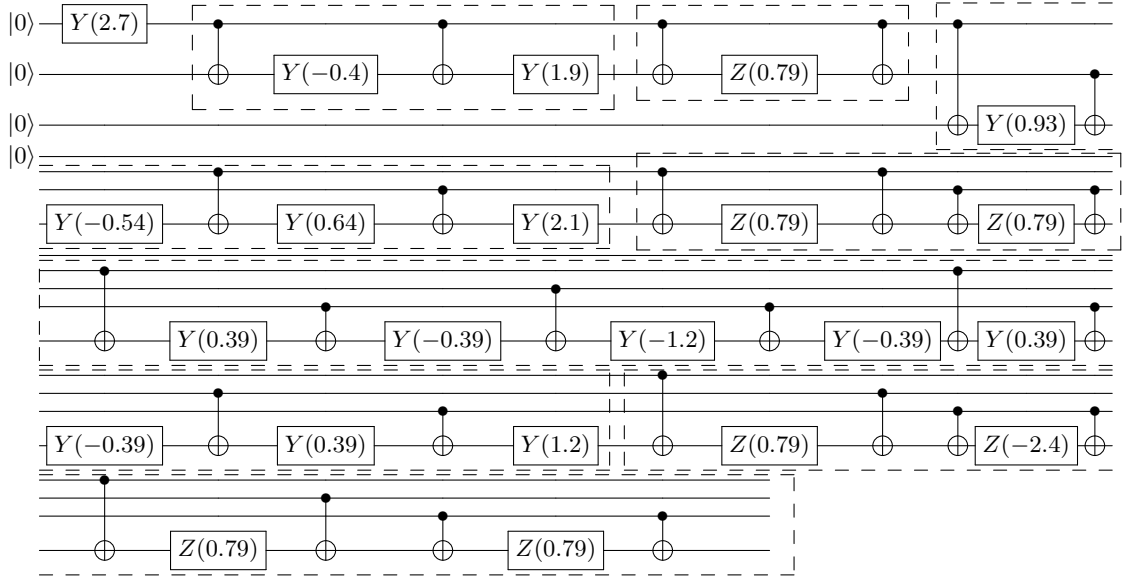
\begin{figure*}[p]
\[
\Qcircuit @C=0.3em @R=.5em  {
	\ket{0}~~	&	&	\gate{Y(2.7)}	&	\ctrl{1}	&	\qw	&	\ctrl{1}	&	\qw	&	\ctrl{1}	&	\qw	&	\ctrl{1}	&	\ctrl{2}	&	\qw	&	\qw	&	\qw\\
\ket{0}~~	&	&	\qw	&	\targ	&	\gate{Y(-0.4)}	&	\targ	&	\gate{Y(1.9)}	&	\targ	&	\gate{Z(0.79)}	&	\targ	&	\qw	&	\qw	&	\ctrl{1}	&	\qw\\
\ket{0}~~	&	&	\qw	&	\qw	&	\qw	&	\qw	&	\qw	&	\qw	&	\qw	&	\qw \ar@{--}[]+<2.2em,-0.8em>;[uu]+<2.2em,0.8em> \ar@{--}[]+<2.2em,5em>;[rrrr]+<0em,5em> \ar@{--}[]+<2.2em,-1em>;[rrrr]+<0em,-1em>	& \targ	&	\gate{Y(0.93)}	&	\targ	&	\qw\\
\ket{0}~~	&	&	\qw	&	\qw	&	\qw	&	\qw	&	\qw	&	\qw	&	\qw	&	\qw	&	\qw	&	\qw	&	\qw	&	\qw\\
		&	&	\qw 	&	\ctrl{2}	&	\qw	&	\qw	&	\qw	&	\ctrl{2}	&	\qw	&	\ctrl{2}	&	\qw	&	\qw	&	\qw	&	\qw\\
	&	&	\qw	&	\qw	&	\qw	&	\ctrl{1}	&	\qw	&	\qw	&	\qw	&	\qw	&	\ctrl{1}	&	\qw	&	\ctrl{1}	&	\qw\\
	&	&	\gate{Y(-0.54)}	&	\targ	&	\gate{Y(0.64)}	&	\targ	&	\gate{Y(2.1)} \ar@{--}[]+<2.2em,-0.8em>;[uu]+<2.2em,0.2em> \ar@{--}[]+<2.2em,2.4em>;[lllll]+<0em,2.4em> \ar@{--}[]+<2.2em,-1em>;[lllll]+<0em,-1em>	&	\targ	&	\gate{Z(0.79)}	&	\targ	&	\targ	&	\gate{Z(0.79)}	&	\targ	&	\qw\\
	&	&	\qw &	\qw	&	\qw	&	\qw	&	\qw	&	\qw	&	\qw	&	\qw	&	\qw	&	\qw	&	\qw	&	\qw\\
	&	&	\ctrl{3}	&	\qw	&	\qw	&	\qw	&	\qw	&	\qw	&	\qw	&	\qw	&	\ctrl{3}	&	\qw	&	\qw	&	\qw\\
	&	&	\qw	&	\qw	&	\qw	&	\qw	&	\ctrl{2}	&	\qw	&	\qw	&	\qw	&	\qw	&	\qw	&	\qw	&	\qw\\
	&	&	\qw	&	\qw	&	\ctrl{1}	&	\qw	&	\qw	&	\qw	&	\ctrl{1}	&	\qw	&	\qw	&	\qw	&	\ctrl{1}	&	\qw\\
	& \ar@{--}[]+<0em,-0.8em>;[uu]+<0em,0.8em> \ar@{--}[]+<0em,3.3em>;[rrrrrrrrrrrr]+<0em,3.3em> \ar@{--}[]+<0em,-1em>;[rrrrrrrrrrrr]+<0em,-1em>	&	\targ	&	\gate{Y(0.39)}	&	\targ	&	\gate{Y(-0.39)}	&	\targ	&	\gate{Y(-1.2)}	&	\targ	&	\gate{Y(-0.39)}	&	\targ	&	\gate{Y(0.39)}	&	\targ	&	\qw\\
	&	&	\qw	&	\qw	&	\qw	&	\qw	&	\qw	&	\ctrl{3}	&	\qw	&	\qw	&	\qw	&	\qw	&	\qw	&	\qw\\
	&	&	\qw	&	\ctrl{2}	&	\qw	&	\qw	&	\qw	&	\qw	&	\qw	&	\ctrl{2}	&	\qw	&	\qw	&	\qw	&	\qw\\
	&	&	\qw	&	\qw	&	\qw	&	\ctrl{1}	&	\qw	&	\qw	&	\qw	&	\qw	&	\ctrl{1}	&	\qw	&	\ctrl{1}	&	\qw\\
	&	&	\gate{Y(-0.39)}	&	\targ	&	\gate{Y(0.39)}	&	\targ	&	\gate{Y(1.2)} \ar@{--}[]+<2.2em,-0.8em>;[uuu]+<2.2em,0.1em> \ar@{--}[]+<2.2em,3.1em>;[lllll]+<0em,3.1em> \ar@{--}[]+<2.2em,-1em>;[lllll]+<0em,-1em>	 
	\ar@{--}[]+<2.8em,-0.8em>;[uuu]+<2.8em,0.1em> \ar@{--}[]+<2.8em,3.1em>;[rrrrrrr]+<0em,3.1em> \ar@{--}[]+<2.8em,-1em>;[rrrrrrr]+<0em,-1em>
	&	\targ 	&	\gate{Z(0.79)}	&	\targ	&	\targ	&	\gate{Z(-2.4)}	&	\targ	&	\qw\\
	&	&	\ctrl{3}	&	\qw	&	\qw	&	\qw	&	\qw	&	\qw	&	\qw\\										
	&	&	\qw	&	\qw	&	\ctrl{2}	&	\qw	&	\qw	&	\qw	&	\qw\\										
	&	&	\qw	&	\qw	&	\qw	&	\ctrl{1}	&	\qw	&	\ctrl{1}	&	\qw\\										
	&	&	\targ	&	\gate{Z(0.79)}	&	\targ	&	\targ	&	\gate{Z(0.79)}	&	\targ	&	\qw \ar@{--}[]+<1em,-0.8em>;[uuu]+<1em,0.1em> \ar@{--}[]+<1em,3.1em>;[lllllll]+<0em,3.1em> \ar@{--}[]+<1em,-1em>;[lllllll]+<0em,-1em> 
	\gategroup{1}{4}{2}{7}{1.3em}{--} \gategroup{1}{8}{2}{10}{1.3em}{--} \gategroup{5}{8}{7}{13}{1.3em}{--} \\  }
\]
\caption{4-qubit circuit which creates a superposition state of the form of Eq. (5) of the main text. This example relies on numerical decomposition methods in terms of quantum multiplexers as described in Ref.~\cite{Shende2006} and is found using Qiskit. Each dashed box represents a single $p$-controlled $R_y$ or $R_z$ multiplexer (where $p$ is necessarily $1\leq p\leq3$). For this random choice of parameters $u_i=(3.64084243, 5.64400577, 4.13234724, 0.73049135, 0.10130389)$, this circuit takes the ground state to the superposition state $\ket{\psi}=0.177936\ket{0011}+0.0161198\ket{0101}+0.156527 \ket{0110}+0.158578\ket{1001}+0.384251\ket{1010}+0.877942\ket{1100}$.}
    \label{fig:multiplexdecomp}
\end{figure*}

Alternatively, if we instead utilize numerical decompositions, then we can significantly reduce the number of required CNOT gates. In this case we use the method described in Ref.~\cite{Shende2006}, which is implemented in Qiskit~\cite{Qiskit}. A similar, but separate method is also utilized in Ref.~\cite{Tubman2018}. This method takes the ground state to an arbitrary $n$-qubit state by use of quantum multiplexers. In a typical VQE algorithm, this numerical method would need to be run at each step of the VQE. However, this algorithm itself is fairly efficient and is computationally much less expensive than the optimization step. An example of this decomposition for a random choice of our state coefficients $u_i$, is shown in Fig.~\ref{fig:multiplexdecomp} and results in, at most, $28$ CNOT gates. In the general case, the decomposition method itself gives a CNOT count of $2^{n+2}-4n-4$ CNOT gates for the generation of a $n$-qubit state. However, Qiskit's transpiler typically reduces the CNOT count of the resulting circuit by an additional factor of 2. The exact CNOT count after transpiliation depends on the target state and therefore fluctuates slightly. We note that this method is not a decomposition of our proposed $E_4$ gate but instead does produce the same output state as the $E_4$ gate when it acts on the ground state.

\begin{figure*}[tb]
\[ \Qcircuit @C=0.75em @R=.3em {
\ket{0} &	&	\gate{X}	&	\ctrl{+1}	&	\qw	&	\ctrl{+1}	&	\qw	&	\ctrl{+1}	&	\targ	&	\qw	&	\ctrlo{+1}	&	\qw	&	\targ	&	\ctrlo{+1}	&	\qw	&	\ctrlo{+1}	&	\qw	&	\ctrlo{+1}	&	\ctrlo{+1}	&	\gate{Y_4^\dagger}	&	\targ	&	\gate{Y_4}	&	\ctrlo{+1}	&	\targ	&	\qw	&	\ctrlo{+1}	&	\qw	&	\targ	&	\qw	\\
\ket{0} &	&	\gate{X}	&	\targ	&	\qw	&	\ctrlo{+1}	&	\qw	&	\targ	&	\ctrlo{-1}	&	\gate{Y_2^\dagger}	&	\targ	&	\gate{Y_2}	&	\ctrlo{-1}	&	\targ	&	\qw	&	\ctrlo{+1}	&	\qw	&	\targ	&	\ctrlo{+1}	&	\qw	&	\ctrlo{-1}	&	\qw	&	\ctrlo{+1}	&	\ctrlo{-1}	&	\gate{Y_5^\dagger}	&	\targ	&	\gate{Y_5}	&	\ctrlo{-1}	&	\qw	\\
\ket{0} &	&	\qw	&	\ctrlo{-1}	&	\gate{Y_1^\dagger}	&	\targ	&	\gate{Y_1}	&	\ctrlo{-1}	&	\ctrl{-1}	&	\qw	&	\ctrl{-1}	&	\qw	&	\ctrl{-1}	&	\ctrl{-1}	&	\qw	&	\ctrl{+1}	&	\qw	&	\ctrl{-1}	&	\targ	&	\qw	&	\ctrlo{-1}	&	\qw	&	\targ	&	\ctrlo{-1}	&	\qw	&	\ctrlo{-1}	&	\qw	&	\ctrlo{-1}	&	\qw	\\
\ket{0} &	&	\qw	&	\ctrlo{-1}	&	\qw	&	\ctrlo{-1}	&	\qw	&	\ctrlo{-1}	&	\ctrlo{-1}	&	\qw	&	\ctrlo{-1}	&	\qw	&	\ctrlo{-1}	&	\ctrlo{-1}	&	\gate{Y_3^\dagger}	&	\targ	&	\gate{Y_3}	&	\ctrlo{-1}	&	\ctrl{-1}	&	\qw	&	\ctrl{-1}	&	\qw	&	\ctrl{-1}	&	\ctrl{-1}	&	\qw	&	\ctrl{-1}	&	\qw	&	\ctrl{-1}	&	\qw	
}
\]
    \caption{Using Gray codes, we show the decomposition of the $E_4$ gate into elementary single and two qubit gates. Decomposition of all Toffoli gates results in a total of 155 CNOT gates and 12 single qubit gates. The notation for the single qubit $Y$ rotations is that $Y_i = R_Y(u_i-\frac{\pi}{2})$.}
    \label{fig:e-gate}
\end{figure*}
\begin{table*}[!p]
  \centering
  \begin{tabular}{@{}cccccc@{}}
\hline
$n,m,s,s_z$            &~~~$N_{\rm T}$~  &~~$N_{\rm C}(E)$~~   &~~$N_{\rm C}(A)$  &~~$N_{\rm C}(N)$ \\ \hline
$n=4,m=2$              &~~~$15$~         &~~$155$~~         &~~$135$  & 28  \\
$n=4,m=2,s=1,s_{z}=0$  &~~~$7$~          &~~$67$~~          &~~$63$   & 14  \\
$n=4,m=2,s=0,s_{z}=0$  &~~~$9$~          &~~$93$~~          &~~$81$  & 24   \\ \hline \hline
$n=6,m=3$              &~~~$57$~         &~~$2337$~~         &~~- & 124   \\
$n=6,m=3,s=\frac{3}{2},s_{z}=\frac{1}{2}$   &~~~$14$~          &~~$574$          &~~- & 62   \\
$n=6,m=3,s=\frac{3}{2},s_{z}=-\frac{1}{2}$  &~~~$14$~          &~~$574$          &~~-    & 62 \\
$n=6,m=3,s=\frac{1}{2},s_{z}=\frac{1}{2}$   &~~~$24$~          &~~$984$         &~~-    & 114 \\
$n=6,m=3,s=\frac{1}{2},s_{z}=-\frac{1}{2}$  &~~~$24$~          &~~$984$         &~~-    & 106 \\ \hline \hline
$n=8,m=4$              &~~~$207$~         &~~$17595$~         &~~-    & 508\\
$n=8,m=4,s=2,s_{z}=1$  &~~~$21$~          &~~$1785$~          &~~-   & 254 \\
$n=8,m=4,s=2,s_{z}=0$  &~~~$35$~          &~~$2975$~          &~~-    & 254\\
$n=8,m=4,s=2,s_{z}=-1$  &~~~$21$~          &~~$1785$~          &~~-   & 254 \\
$n=8,m=4,s=1,s_{z}=1$  &~~~$45$~          &~~$3825$~          &~~-   & 454 \\
$n=8,m=4,s=1,s_{z}=0$  &~~~$71$~          &~~$6885$~          &~~-   & 508 \\
$n=8,m=4,s=1,s_{z}=-1$  &~~~$45$~          &~~$3825$~          &~~-   & 454 \\
$n=8,m=4,s=0,s_{z}=0$  &~~~$77$~          &~~$6545$~          &~~-   & 432 \\ \hline
\end{tabular}
\caption{
For each spin configuration, we count the total number of Toffoli gates, $N_{\rm T}$, and CNOT gates required in each respective circuit using our $E$ gate construction. $N_{\rm C}(E)$ represents exact Toffoli decomposition while $N_{\rm C}(A)$ represents approximate decomposition. When $n \ge 6$, $N_{\rm C}(E) = N_{\rm T} (2n^{2}-6n+5)$~\cite{szyprowski2013}. We include $N_{\rm C}(N)$, which shows the maximum possible CNOT count using the numerical methods in Ref.~\cite{Shende2006} and after transpiling in Qiskit, for any random target state in the chosen subspace.}
\label{Number_Table}
\end{table*}
Next, we consider the same problem but now for the case of three fermions in six spin-orbitals. The different spin subspaces are now given by
\begin{equation}
\ket{s=\frac{3}{2},s_z=+\frac{1}{2}}=\frac{1}{\sqrt{3}}(\ket{011010}+\ket{100110}+\ket{101001}),\nonumber
\end{equation}
\begin{equation}
\ket{s=\frac{3}{2},s_z=-\frac{1}{2}}=\frac{1}{\sqrt{3}}(\ket{010110}+\ket{011001}+\ket{100101}),\nonumber
\end{equation}
\begin{equation}
    \begin{split}
\ket{s=\frac{1}{2},s_z=+\frac{1}{2}}=&\alpha\ket{001110}+\beta\ket{001011}+\gamma\ket{110010}\\
+&\delta\ket{111000}+\epsilon\ket{100011}+\eta\ket{101100}\\
+&\frac{\chi}{\sqrt{2}}(\ket{101001}-\ket{100110})\\
+&\frac{\xi}{\sqrt{6}}(\ket{100110}+\ket{101001}-2\ket{011010}),\nonumber
\end{split}
\end{equation}
\begin{equation}
\begin{split}
\ket{s=\frac{1}{2},s_z=-\frac{1}{2}}=&\alpha\ket{010011}+\beta\ket{011100}+\gamma\ket{000111}\\
+&\delta\ket{110100}+\epsilon\ket{001101}+\eta\ket{110001}\\
+&\frac{\chi}{\sqrt{2}}(\ket{011001}-\ket{010110})\\
+&\frac{\xi}{\sqrt{6}}(2\ket{100101}-\ket{010110}-\ket{011001}).\nonumber
\end{split}
\end{equation}
If we construct an $E_6$ unitary analogous to Eq.~\eqref{eq:E4gate} and perform a similar decomposition in terms of Toffoli gates, we obtain the circuit shown in Fig.~\ref{fig:f-gate}. A summary of gate counts for $E_4$, $E_6$, and $E_8$ is given in Table~\ref{Number_Table}.

If we again consider the numerical decomposition methods that construct the desired superposition state, then we can construct such a state in only $124$ CNOT gates. Using Qiskit's transpiler, the CNOT   count for constructing an $n$-qubit state ($n\in 2\mathbb{N}$) with only $n/2$ particles is, at most, $2^{n+1}-4$ CNOT's.

\begin{figure*}[tb]
\[ \Qcircuit @C=0.75em @R=0.3em {
\ket{0} &	&	\gate{X}	&	\ctrl{+1}	&	\qw	&	\ctrl{+1}	&	\qw	&	\ctrl{+1}	&	\ctrl{+1}	&	\qw	&	\ctrl{+1}	&	\qw	&	\ctrl{+1}	&	\ctrl{+1}	&	\qw	&	\ctrl{+1}	&	\qw	&	\ctrl{+1}	&	\ctrl{+1}	&	\qw	&	\ctrl{+1}	&	\qw	&	\qw	\\
\ket{0} &	&	\gate{X}	&	\ctrl{+1}	&	\qw	&	\ctrl{+1}	&	\qw	&	\ctrl{+1}	&	\ctrl{+1}	&	\qw	&	\ctrl{+1}	&	\qw	&	\ctrl{+1}	&	\ctrl{+1}	&	\qw	&	\ctrl{+1}	&	\qw	&	\ctrl{+1}	&	\targ	&	\qw	&	\ctrlo{+1}	&	\qw	&	\qw	\\
\ket{0} &	&	\gate{X}	&	\targ	&	\qw	&	\ctrlo{+1}	&	\qw	&	\targ	&	\ctrlo{+1}	&	\qw	&	\ctrlo{+1}	&	\qw	&	\ctrlo{+1}	&	\ctrlo{+1}	&	\qw	&	\ctrlo{+1}	&	\qw	&	\ctrlo{+1}	&	\ctrlo{-1}	&	\gate{Y_4^\dagger}	&	\targ	&	\gate{Y_4}	&	\qw	\\
\ket{0} &	&	\qw	&	\ctrlo{-1}	&	\gate{Y_1^\dagger}	&	\targ	&	\gate{Y_1}	&	\ctrlo{-1}	&	\targ	&	\qw	&	\ctrlo{+1}	&	\qw	&	\targ	&	\ctrlo{+1}	&	\qw	&	\ctrlo{+1}	&	\qw	&	\ctrlo{+1}	&	\ctrlo{-1}	&	\qw	&	\ctrlo{-1}	&	\qw	&	\qw	\\
\ket{0} &	&	\qw	&	\ctrlo{-1}	&	\qw	&	\ctrlo{-1}	&	\qw	&	\ctrlo{-1}	&	\ctrlo{-1}	&	\gate{Y_2^\dagger}	&	\targ	&	\gate{Y_2}	&	\ctrlo{-1}	&	\targ	&	\qw	&	\ctrlo{+1}	&	\qw	&	\targ	&	\ctrlo{-1}	&	\qw	&	\ctrlo{-1}	&	\qw	&	\qw	\\
\ket{0} &	&	\qw	&	\ctrlo{-1}	&	\qw	&	\ctrlo{-1}	&	\qw	&	\ctrlo{-1}	&	\ctrlo{-1}	&	\qw	&	\ctrlo{-1}	&	\qw	&	\ctrlo{-1}	&	\ctrlo{-1}	&	\gate{Y_3^\dagger}	&	\targ	&	\gate{Y_3}	&	\ctrlo{-1}	&	\ctrl{-1}	&	\qw	&	\ctrl{-1}	&	\qw	&	\qw	\\
&	&	\ctrl{+1}	&	\ctrl{+1}	&	\qw	&	\ctrl{+1}	&	\qw	&	\ctrl{+1}	&	\ctrl{+1}	&	\qw	&	\ctrl{+1}	&	\qw	&	\ctrl{+1}	&	\ctrl{+1}	&	\qw	&	\ctrl{+1}	&	\qw	&	\ctrl{+1}	&	\ctrl{+1}	&	\qw	&	\ctrl{+1}	&	\qw	&	\qw	\\
&	&	\targ	&	\ctrlo{+1}	&	\qw	&	\ctrlo{+1}	&	\qw	&	\ctrlo{+1}	&	\ctrlo{+1}	&	\qw	&	\ctrlo{+1}	&	\qw	&	\ctrlo{+1}	&	\ctrlo{+1}	&	\qw	&	\ctrlo{+1}	&	\qw	&	\ctrlo{+1}	&	\ctrlo{+1}	&	\qw	&	\ctrlo{+1}	&	\qw	&	\qw	\\
&	&	\ctrlo{-1}	&	\ctrl{+1}	&	\qw	&	\ctrl{+1}	&	\qw	&	\ctrl{+1}	&	\ctrl{+1}	&	\qw	&	\ctrl{+1}	&	\qw	&	\ctrl{+1}	&	\targ	&	\qw	&	\ctrlo{+1}	&	\qw	&	\targ	&	\ctrlo{+1}	&	\qw	&	\ctrlo{+1}	&	\qw	&	\qw	\\
&	&	\ctrlo{-1}	&	\ctrlo{+1}	&	\gate{Y_5^\dagger}	&	\targ	&	\gate{Y_5}	&	\ctrlo{+1}	&	\targ	&	\qw	&	\ctrlo{+1}	&	\qw	&	\targ	&	\ctrlo{-1}	&	\gate{Y_7^\dagger}	&	\targ	&	\gate{Y_7}	&	\ctrlo{-1}	&	\ctrl{+1}	&	\qw	&	\ctrl{+1}	&	\qw	&	\qw	\\
&	&	\ctrlo{-1}	&	\ctrlo{+1}	&	\qw	&	\ctrlo{-1}	&	\qw	&	\ctrlo{+1}	&	\ctrlo{-1}	&	\gate{Y_6^\dagger}	&	\targ	&	\gate{Y_6}	&	\ctrlo{-1}	&	\ctrl{-1}	&	\qw	&	\ctrl{-1}	&	\qw	&	\ctrl{-1}	&	\targ	&	\qw	&	\ctrlo{+1}	&	\qw	&	\qw	\\
&	&	\ctrl{-1}	&	\targ	&	\qw	&	\ctrlo{-1}	&	\qw	&	\targ	&	\ctrlo{-1}	&	\qw	&	\ctrlo{-1}	&	\qw	&	\ctrlo{-1}	&	\ctrlo{-1}	&	\qw	&	\ctrlo{-1}	&	\qw	&	\ctrlo{-1}	&	\ctrlo{-1}	&	\gate{Y_8^\dagger}	&	\targ	&	\gate{Y_8}	&	\qw	\\
&	&	\ctrl{+1}	&	\ctrl{+1}	&	\qw	&	\ctrl{+1}	&	\qw	&	\ctrl{+1}	&	\targ	&	\qw	&	\ctrlo{+1}	&	\qw	&	\targ	&	\ctrlo{+1}	&	\qw	&	\ctrlo{+1}	&	\qw	&	\ctrlo{+1}	&	\ctrlo{+1}	&	\qw	&	\ctrlo{+1}	&	\qw	&	\qw	\\
&	&	\ctrlo{+1}	&	\ctrlo{+1}	&	\qw	&	\ctrlo{+1}	&	\qw	&	\ctrlo{+1}	&	\ctrlo{-1}	&	\gate{Y_{10}^\dagger}	&	\targ	&	\gate{Y_{10}}	&	\ctrlo{-1}	&	\ctrl{+1}	&	\qw	&	\ctrl{+1}	&	\qw	&	\ctrl{+1}	&	\ctrl{+1}	&	\qw	&	\ctrl{+1}	&	\qw	&	\qw	\\
&	&	\ctrlo{+1}	&	\ctrlo{+1}	&	\qw	&	\ctrlo{+1}	&	\qw	&	\ctrlo{+1}	&	\ctrlo{-1}	&	\qw	&	\ctrlo{-1}	&	\qw	&	\ctrlo{-1}	&	\ctrlo{+1}	&	\qw	&	\ctrlo{+1}	&	\qw	&	\ctrlo{+1}	&	\ctrlo{+1}	&	\qw	&	\ctrlo{+1}	&	\qw	&	\qw	\\
&	&	\ctrl{+1}	&	\targ	&	\qw	&	\ctrlo{+1}	&	\qw	&	\targ	&	\ctrlo{-1}	&	\qw	&	\ctrlo{-1}	&	\qw	&	\ctrlo{-1}	&	\ctrlo{+1}	&	\gate{Y_{11}^\dagger}	&	\targ	&	\gate{Y_{11}}	&	\ctrlo{+1}	&	\ctrl{+1}	&	\qw	&	\ctrl{+1}	&	\qw	&	\qw	\\
&	&	\targ	&	\ctrlo{-1}	&	\gate{Y_9^\dagger}	&	\targ	&	\gate{Y_9}	&	\ctrlo{-1}	&	\ctrl{-1}	&	\qw	&	\ctrl{-1}	&	\qw	&	\ctrl{-1}	&	\targ	&	\qw	&	\ctrlo{-1}	&	\qw	&	\targ	&	\ctrlo{+1}	&	\gate{Y_{12}^\dagger}	&	\targ	&	\gate{Y_{12}}	&	\qw	\\
&	&	\ctrlo{-1}	&	\ctrl{-1}	&	\qw	&	\ctrl{-1}	&	\qw	&	\ctrl{-1}	&	\ctrl{-1}	&	\qw	&	\ctrl{-1}	&	\qw	&	\ctrl{-1}	&	\ctrl{-1}	&	\qw	&	\ctrl{-1}	&	\qw	&	\ctrl{-1}	&	\targ	&	\qw	&	\ctrlo{-1}	&	\qw	&	\qw	\\
&	&	\ctrlo{+1}	&	\ctrlo{+1}	&	\qw	&	\ctrlo{+1}	&	\qw	&	\ctrlo{+1}	&	\ctrlo{+1}	&	\qw	&	\ctrlo{+1}	&	\qw	&	\ctrlo{+1}	&	\ctrlo{+1}	&	\qw	&	\ctrlo{+1}	&	\qw	&	\ctrlo{+1}	&	\ctrlo{+1}	&	\qw	&	\ctrlo{+1}	&	\qw	&	\qw	\\
&	&	\ctrl{+1}	&	\ctrl{+1}	&	\qw	&	\ctrl{+1}	&	\qw	&	\ctrl{+1}	&	\ctrl{+1}	&	\qw	&	\ctrl{+1}	&	\qw	&	\ctrl{+1}	&	\ctrl{+1}	&	\qw	&	\ctrl{+1}	&	\qw	&	\ctrl{+1}	&	\targ	&	\qw	&	\ctrlo{+1}	&	\qw	&	\qw	\\
&	&	\ctrlo{+1}	&	\ctrlo{+1}	&	\gate{Y_{13}^\dagger}	&	\targ	&	\gate{Y_{13}}	&	\ctrlo{+1}	&	\ctrl{+1}	&	\qw	&	\ctrl{+1}	&	\qw	&	\ctrl{+1}	&	\ctrl{+1}	&	\qw	&	\ctrl{+1}	&	\qw	&	\ctrl{+1}	&	\ctrl{-1}	&	\qw	&	\ctrl{+1}	&	\qw	&	\qw	\\
&	&	\ctrl{+1}	&	\ctrl{+1}	&	\qw	&	\ctrl{-1}	&	\qw	&	\ctrl{+1}	&	\targ	&	\qw	&	\ctrlo{+1}	&	\qw	&	\targ	&	\ctrlo{+1}	&	\qw	&	\ctrlo{+1}	&	\qw	&	\ctrlo{+1}	&	\ctrlo{-1}	&	\gate{Y_{16}^\dagger}	&	\targ	&	\gate{Y_{16}}	&	\qw	\\
&	&	\ctrlo{+1}	&	\targ	&	\qw	&	\ctrlo{-1}	&	\qw	&	\targ	&	\ctrlo{-1}	&	\gate{Y_{14}^\dagger}	&	\targ	&	\gate{Y_{14}}	&	\ctrlo{-1}	&	\targ	&	\qw	&	\ctrlo{+1}	&	\qw	&	\targ	&	\ctrlo{-1}	&	\qw	&	\ctrlo{-1}	&	\qw	&	\qw	\\
&	&	\targ	&	\ctrlo{-1}	&	\qw	&	\ctrlo{-1}	&	\qw	&	\ctrlo{-1}	&	\ctrlo{-1}	&	\qw	&	\ctrlo{-1}	&	\qw	&	\ctrlo{-1}	&	\ctrlo{-1}	&	\gate{Y_{15}^\dagger}	&	\targ	&	\gate{Y_{15}}	&	\ctrlo{-1}	&	\ctrl{-1}	&	\qw	&	\ctrl{-1}	&	\qw	&	\qw	\\
&	&	\ctrlo{+1}	&	\ctrlo{+1}	&	\qw	&	\ctrlo{+1}	&	\qw	&	\ctrlo{+1}	&	\ctrlo{+1}	&	\qw	&	\ctrlo{+1}	&	\qw	&	\ctrlo{+1}	&	\ctrlo{+1}	&	\qw	&	\ctrlo{+1}	&	\qw	&	\ctrlo{+1}	&	\qw	&	\qw	&	\qw	&	\qw	&	\qw	\\
&	&	\targ	&	\ctrlo{+1}	&	\qw	&	\ctrlo{+1}	&	\qw	&	\ctrlo{+1}	&	\ctrlo{+1}	&	\qw	&	\ctrlo{+1}	&	\qw	&	\ctrlo{+1}	&	\ctrlo{+1}	&	\qw	&	\ctrlo{+1}	&	\qw	&	\ctrlo{+1}	&	\qw	&	\qw	&	\qw	&	\qw	&	\qw	\\
&	&	\ctrl{-1}	&	\ctrl{+1}	&	\qw	&	\ctrl{+1}	&	\qw	&	\ctrl{+1}	&	\ctrl{+1}	&	\qw	&	\ctrl{+1}	&	\qw	&	\ctrl{+1}	&	\targ	&	\qw	&	\ctrlo{+1}	&	\qw	&	\targ	&	\qw	&	\qw	&	\qw	&	\qw	&	\qw	\\
&	&	\ctrlo{-1}	&	\ctrl{+1}	&	\qw	&	\ctrl{+1}	&	\qw	&	\ctrl{+1}	&	\targ	&	\qw	&	\ctrlo{+1}	&	\qw	&	\targ	&	\ctrlo{-1}	&	\gate{Y_{19}^\dagger}	&	\targ	&	\gate{Y_{19}}	&	\ctrlo{-1}	&	\qw	&	\qw	&	\qw	&	\qw	&	\qw	\\
&	&	\ctrlo{-1}	&	\ctrlo{+1}	&	\gate{Y_{17}^\dagger}	&	\targ	&	\gate{Y_{17}}	&	\ctrlo{+1}	&	\ctrl{-1}	&	\qw	&	\ctrl{+1}	&	\qw	&	\ctrl{-1}	&	\ctrl{-1}	&	\qw	&	\ctrl{-1}	&	\qw	&	\ctrl{-1}	&	\qw	&	\qw	&	\qw	&	\qw	&	\qw	\\
&	&	\ctrl{-1}	&	\targ	&	\qw	&	\ctrlo{-1}	&	\qw	&	\targ	&	\ctrlo{-1}	&	\gate{Y_{18}^\dagger}	&	\targ	&	\gate{Y_{18}}	&	\ctrlo{-1}	&	\ctrl{-1}	&	\qw	&	\ctrl{-1}	&	\qw	&	\ctrl{-1}	&	\qw	&	\qw	&	\qw	&	\qw	&	\qw	
}
\]
    \caption{Using Gray codes, we show the decomposition of the $E_6$ gate into elementary single and two qubit gates. Decomposition of all Toffoli gates results in a total of 2337 CNOT gates and 41 single qubit gates. The notation for the single qubit $Y$ rotations is that $Y_i = R_Y(u_i-\frac{\pi}{2})$.}
    \label{fig:f-gate}
\end{figure*}

\subsection{Numerical Verification of Circuits}
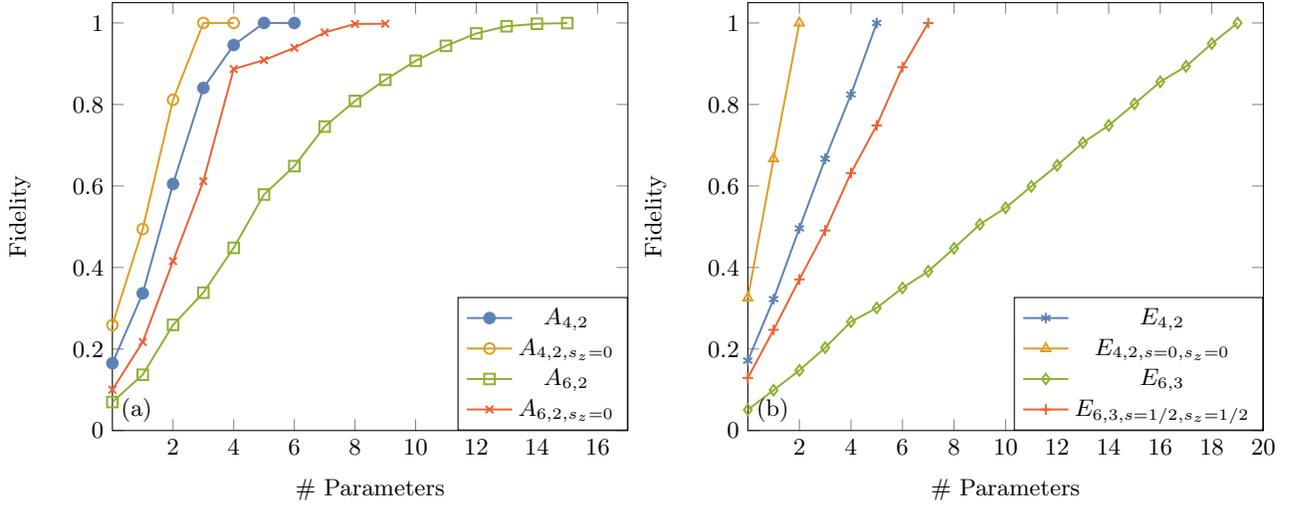
\begin{figure*}[!htb]
\centering
    \definecolor{mmaBlue}{HTML}{5e81b5}
    \definecolor{mmaOrange}{HTML}{e19c24}
    \definecolor{mmaGreen}{HTML}{8fb032}
    \definecolor{mmaRed}{HTML}{eb6235}
    \definecolor{mmaPurple}{HTML}{8778b3}
    \definecolor{mmaBrown}{HTML}{c56e1a}
    
    \begin{tikzpicture}
          \begin{axis}[
            xlabel = {\# Parameters},
            ylabel = {Fidelity},
            xmin = 0,
            xmax = 17,
            ymin=0,
            ymax=1.05,
            xtick = {2,4,6,8,10,12,14,16},
            legend style={at={(1,0)},anchor=south east}
            ]
    \addlegendentry{$A_{4,2}$}
    \addlegendentry{$A_{4,2,s_z=0}$}
    \addlegendentry{$A_{6,2}$}
    \addlegendentry{$A_{6,2,s_z=0}$}
            \addplot[line width=0.75pt,solid,color=mmaBlue,mark=*]
                table{
                  0 0.16509149364639392
                  1 0.33665252551989205 
                  2 0.6049371516211911
                  3 0.8404554393468447
                  4 0.9459059543301387
                  5 0.9999999999999871
                  6 0.9999999999998542
                   };
            \addplot[line width=0.75pt,solid,color=mmaOrange,mark=o]
                table{
                    0 0.25878293116520046
                    1 0.49440392074839595
                    2 0.8115117995129308
                    3 0.9999999999999991
                    4 0.9999999999999994
                    };
            \addplot[line width=0.75pt,solid,color=mmaGreen,mark=square]
                table{
                    0 0.06992015337734175
                    1 0.13686939535030176
                    2 0.25920784546740644
                    3 0.33821724072512843
                    4 0.4482138636307883
                    5 0.5790189972629262 
                    6 0.6487788429764421
                    7 0.7457698877589277
                    8 0.8086253353703974
                    9 0.860500576858853
                    10 0.9071898143147447
                    11 0.9438958013481336
                    12 0.9742950314763058
                    13 0.9918899837042585
                    14 0.9980404361463416
                    15 0.9997474639172255
                };
                \addplot[line width=0.75pt,solid,color=mmaRed,mark=x]
                table{
                    0 0.10014059991729667
                    1 0.21724807938940963
                    2 0.4157067284134691
                    3 0.6117889916930895
                    4 0.8869307355074671
                    5 0.908927035628553
                    6 0.9389732646218656
                    7 0.9767104030228243
                    8 0.9975587021887726
                    9 0.9981579125185256
                }; 
        \end{axis}
        \node [above right]  {(a)};
    \end{tikzpicture}
   \begin{tikzpicture}
        \begin{axis}[
            xlabel = {\# Parameters},
            ylabel = {Fidelity},
            xmin = 0,
            xmax = 20,
            ymin=0,
            ymax=1.05,
            xtick = {2,4,6,8,10,12,14,16,18,20},
            legend style={at={(1,0)},anchor=south east}
            ]
    \addlegendentry{$E_{4,2}$}
    \addlegendentry{$E_{4,2,s=0,s_z=0}$}
    \addlegendentry{$E_{6,3}$}
    \addlegendentry{$E_{6,3,s=1/2,s_z=1/2}$}
                \addplot[line width=0.75pt,solid,color=mmaBlue,mark=asterisk]
                table{
                   0 0.1715653067999831
                   1 0.32247880115410454
                   2 0.4960399942562949
                   3 0.6666999523418737
                   4 0.8244949075978281
                   5 1
                };
                \addplot[line width=0.75pt,solid,color=mmaOrange,mark=triangle]
                table{
                   0 0.32548269862318585
                   1 0.6669347865633025
                   2 1
                };
                \addplot[line width=0.75pt,solid,color=mmaGreen,mark=diamond]
                table{
                    0 0.05103360413481167
                    1 0.09909949570780162
                    2 0.14763594555485882
                    3 0.2034360771691898
                    4 0.2669254793226245
                    5 0.3008989444203076
                    6 0.3498767949191538
                    7 0.3906850623433296
                    8 0.4473183219217465
                    9 0.5058434757034792
                    10 0.5464702122461437
                    11 0.5991913248879086
                    12 0.6505214763460314
                    13 0.706213523203654
                    14 0.748580501381204
                    15 0.8016026083183914
                    16 0.855659111208211
                    17 0.8935938125364372
                    18 0.9494158734712603
                    19 0.9999999999999989
                };
                \addplot[line width=0.75pt,solid,color=mmaRed,mark=+]
                table{
                    0 0.12913804148455005
                    1 0.2471502933637832
                    2 0.370772705929965
                    3 0.4907368501202262
                    4 0.6318245547097416
                    5 0.7487032017691649
                    6 0.8915439777152843
                    7 0.999972
                };
                \node [above right]  {(b)};
                \end{axis}
        \end{tikzpicture}
\caption{Fidelity $F$ (defined in the main text) of our state preparation circuits as a function of the number of variational circuit parameters. (a) Fidelity of our $A_{n,m}$ circuits for $n$ qubits and $m$ excitations (particles). For $n=4,m=2$ and spin projection left unspecified, these gates can achieve unit fidelity with only the minimal $\binom{4}{2}-1=5$ parameters (blue solid circles). Restricting to states with a spin projection eigenvalue of $s_z=0$ further reduces the number of minimal parameters to three (orange open circles). For the case $n=6,m=2$, a similar circuit achieves unit fidelity with the minimal number of parameters $\binom{6}{2}-1=14$ (green squares). Again restricting to the $s_z=0$ subspace reduces the number of minimal parameters to eight (red x's). One extra parameter is included in each case to show that additional gates are unnecessary. (b) Fidelity of our $E_{n,m}$ gates and total-spin-conserving $E_{n,m,s,s_z}$ gates. These gates also achieve unit fidelity using the minimal number of parameters needed to span the relevant Hilbert subspace.}
\label{fig:fidparams}
\end{figure*}

In order to confirm the validity of our $A$ gate circuits, we compute the circuit fidelity as a function of the number of variational parameters for several example cases. We first consider one of our core examples from the main text, Fig. 3, which shows an $A$ gate circuit for $m=2$ electrons in $n=4$ orbitals. A general state in this subspace can be written as 
\begin{equation}
\alpha \ket{1100}+\beta\ket{1010}+\gamma\ket{1001}+\delta\ket{0110}+\eta\ket{0101}+\chi\ket{0011},
\label{eq:gen4choose2}
\end{equation} which involves the $\binom{4}{2}=6$ basis states that span the number conserving subspace. Following the discussion in the main text, our proposed circuit should span this space using the fewest possible parameters. For clarity, we follow the suggestions in the main text for time-reversal symmetry and therefore set all $\phi_i$ to zero and fix $\theta_5$ to zero. With these settings, this circuit then produces the following parameterized state, when it acts on the ground state $\ket{0000}$,
\begin{equation}
\begin{split}
&\ket{0110}(-C_2 C_3 S_1 S_4-C_1 C_2 C_4)\\
+&\ket{1001}(C_4 S_1 S_2+C_1 C_3 S_4 S_2)\\
+&\ket{1010}(C_2 C_3 C_4 C_6 S_1+C_2 S_3 S_6 S_1-C_1 C_2 C_6 S_4)\\
+&\ket{0101}(-C_1 C_3 C_4 C_6 S_2+C_6 S_1 S_4 S_2-C_1 S_3 S_6 S_2)\\
+&\ket{1100}(-C_2 C_6 S_1 S_3+C_2 C_3 C_4 S_1 S_6-C_1 C_2 S_4 S_6)\\
+&\ket{0011}(-C_1 C_6 S_2 S_3+C_1 C_3 C_4 S_2 S_6-S_1 S_2 S_4 S_6),
\end{split}
\label{eq:a42}
\end{equation}
where $C_i=\cos(\theta_i),S_i=\sin(\theta_i)$. From here we can now sample over random choices of coefficients in Eq.~\ref{eq:gen4choose2} and calculate the fidelity, $F=\frac{1}{N}\sum_{i=1}^{N}|\langle  \Psi(\theta_j)|\psi_i\rangle|^2$, maximizing over $\theta_j$. We show in Fig.~\ref{fig:fidparams} that the fidelity of this circuit increases monotonically with the number of $\theta_j$ parameters and saturates at unity once this number equals the dimension of the symmetry subspace, which is 5 in this case. The fidelity is computed by averaging over $N=2000$ random states. We also performed similar numerical checks for the other choices of number of qubits ($n$) and the number of excitations ($m$), including all ($n>m$) permutations of $n=\{2,3,4,5,6\},m=\{1,2,3,4,5\}$, following the general construction outlined in the main text. The case $n=6$, $m=2$ is also included in the figure (green curve), where it is again evident that the fidelity reaches unity when the number of parameters equals the subspace dimension. The same behavior arose in all the examples we checked.

\begin{figure}[!tb]
\[ \Qcircuit @C=0.5em @R=.7em {
\ket{0} &	&	\gate{X}	&	\multigate{1}{A(\theta_1,0)}	&	\qw	&	\multigate{1}{A(\theta_4,0)}	&	\qw	&	\qw	\\
\ket{0} &	&	\qw	&	\ghost{A(\theta_1,0)}	&	\multigate{1}{A(0,0)}	&	\ghost{A(\theta_4,0)}	&	\qw	&	\qw	\\
\ket{0} &	&	\qw	&	\multigate{1}{A(\theta_2,0)}	&	\ghost{A(0,0)}	&	\qw	&	\qw	&	\qw	\\
\ket{0} &	&	\gate{X}	&	\ghost{A(\theta_2,0)}	&	\qw	&	\qw	&	\qw	&	\qw
} \]
\caption{An example circuit for the case of time reversal symmetry with $n=4,m=2,s_z=0$ which exactly spans the subspace defined by four basis states using the minimal number (3) of parameters.}
\label{fig:a42sz}
\end{figure}
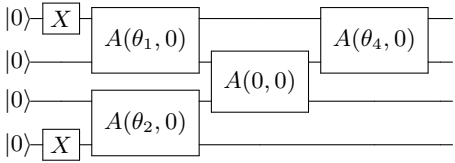

Next we discuss the enforcement of $s_z$ symmetry using this same circuit. In order to take advantage of $s_z$ symmetry, we only require slight modifications to the base circuit as described in the main text. First, we need to assign spin labels to the qubits. We choose a block assignment for this example, where all spin-up orbitals are assigned to the top half and all spin-down orbitals are assigned to the bottom half of the qubits. If we consider the case of $n=4,m=2,s_z=0$, the states that span this spin subspace are
\[
\beta\ket{1010}+\gamma\ket{1001}+\delta\ket{0110}+\eta\ket{0101},
\]
which is just a reduced version of Eq.~\ref{eq:gen4choose2}, due to our $s_z=0$ restriction.
With this choice, it is clear that if we start with an initial state with a quantum number of $s_z=0$ and do not mix the split spin subspaces, then the circuit will preserve this symmetry. In order to illustrate this, we include the modification of the original circuit to maintain this symmetry (also assuming time-reversal symmetry), in Fig.~\ref{fig:a42sz}. Here, we can see that we have removed some $A$ gates from the general circuit case ($n=4, m=2$, without any spin symmetry) in order to reduce the parameter count to the dimension of this reduced space, following the protocol in the main text. Similar to the previous example, this modified circuit produces the following state when it acts on the ground state,
\begin{equation}
\begin{split}
&C_1 S_{2-3}\ket{1010}+C_1C_{2-3}\ket{1001}\\
&+S_1 S_{2+3}\ket{0110}+S_1 C_{2+3}\ket{0101},
\end{split}
\end{equation}
where $C_{i\pm j}$ abbreviates for $\cos(\theta_1\pm \theta_2)$.
In this case, we only require 3 parameters to span the relevant subspace and again achieve unit fidelity in this case as shown in Fig.~\ref{fig:fidparams}(a) (yellow curve). The case $n=6$, $m=2$ is also shown in the figure (red curve).

In all cases shown in Fig.~\ref{fig:fidparams}(a), we find the optimal point where we achieve unit fidelities(with minimal parameters for the chosen space) and fix parameters to zero (starting from the ``end" of the circuit) to show how the fidelity decreases when too few parameters are used. For our $A$ gate circuits, over-parameterization is also indicated by including extra parameters (inserting extra gates), which of course maintains the fidelity at unity. Although numerical verification is not necessary in the case of our $E$ gate circuits, which conserve total spin in addition to particle number, spin magnetization, and time-reversal symmetries, we also show the fidelities of these circuits in Fig. S4(b) for comparison. As expected, the fidelity again saturates at unity when the number of parameters reaches the dimenions of the symmetry subspace. Our $E_4,E_6$ gates show an approximate linear increase in subspace coverage with increasing parameters. We include these examples cases in an skeleton notebook on Github~\cite{Gardgit2019}.

\raggedright